\documentclass[aps,prc,twocolumn,10pt, superscriptaddress, showpacs, floatfix]{revtex4-1}

\usepackage{amssymb,epsfig}
\usepackage{bbold}

\newcommand{\be}{\begin{equation}}
\newcommand{\ee}{\end{equation}}
\newcommand{\bea}{\begin{eqnarray}}
\newcommand{\eea}{\end{eqnarray}}

\begin{document}

\title{On the dynamical kernels of fermionic equations of motion in strongly-correlated media}
\author{Elena Litvinova}
\affiliation{Department of Physics, Western Michigan University, Kalamazoo, MI 49008, USA}
\affiliation{Facility for Rare Isotope Beams, Michigan State University, East Lansing, MI 48824, USA}
\affiliation{GANIL, CEA/DRF-CNRS/IN2P3, F-14076 Caen, France}

\date{\today}

\begin{abstract}
Two-point fermionic propagators in strongly correlated media are considered with an emphasis on the dynamical interaction kernels of their equations of motion (EOM). With the many-body Hamiltonian confined by a two-body interaction, the EOMs for the two-point fermionic propagators acquire the Dyson form and, before taking any approximation, the interaction kernels decompose into the static and dynamical (time-dependent) contributions. The latter translate to the energy-dependent and the former map to the energy-independent terms in the energy domain. We dwell particularly on the energy-dependent terms, which generate long-range correlations while making feedback on their short-range static counterparts. The origin, forms, and various approximations for the dynamical kernels of one-fermion and two-fermion propagators, most relevant in the intermediate-coupling regime, are discussed. Applications to the electromagnetic dipole response of $^{68,70}$Ni and low-energy quadrupole response of $^{114,116,124}$Sn are presented.  
\end{abstract}
\maketitle

\section{Introduction} 

Finding accurate quantitative solutions to the nuclear many-body problem has remained an active field of research for decades. Challenged by the requirements of various applications in nuclear science and technology and driven by the newly emerging computational capabilities, it shows substantial progress over the years, however, predictive computation of atomic nuclei calls for further developments on the theoretical side. 

One of the most powerful tools to study atomic nuclei and, more generally, strongly correlated fermionic many-body systems is the Green function method. Various Green functions, or propagators, belonging to a larger class of correlation functions (CFs), form a common theoretical background across the energy scales. The Green functions are directly related to well-defined observables \cite{Matsubara1955,Watson1956,Brueckner1955,Brueckner1955a,Martin1959,Ethofer1969,SchuckEthofer1973}. In particular, the single-nucleon propagators are linked to the energies of odd-particle systems and spectroscopic factors which, in the case of atomic nuclei, can be extracted from, e.g., transfer or knock-out reactions. The two-nucleon particle-hole propagators are associated with the response to external probes of electromagnetic, strong, or weak character. Superfluidity can be efficiently described by two-particle in-medium pair propagators associated with pair transfer, while the residues of those propagators can be related to the pairing gaps \cite{Gorkov1958,Kadanoff1961}. 

While these propagators are mostly relevant to the observed phenomena, higher-rank CFs appear as part of the general theory in the dynamical kernels of the equations of motion (EOMs) describing the lower-rank ones \cite{Martin1959,Ethofer1969,EthoferSchuck1969,SchuckEthofer1973}. 
The higher-rank propagators quantify the dynamical in-medium effects of long-range correlations thus promoting the connections between the degrees of freedom across the energy scales. They are the source of coupling between EOMs for propagators of all ranks allowable in the given system into a hierarchy, which can be decoupled by making certain approximations. 
Note here that the famous EOM method of Rowe \cite{Rowe1968} operating directly bosonic excitation operators shows similar features and can be viewed as a counterpart to the Green function method. 
Besides the perturbative treatment, useful mostly in weak-coupling limit, the EOMs can be decoupled by cluster decompositions of their dynamical kernels in either non-symmetric \cite{Gorkov1958,Martin1959} or symmetric \cite{Schuck1976,AdachiSchuck1989,Danielewicz1994,DukelskyRoepkeSchuck1998} forms, with varied correlation content. 

Certain advantages of the symmetric forms of the dynamical kernels in the context of their cluster decompositions were particularly pointed out and elaborated by Peter Schuck and coauthors \cite{AdachiSchuck1989,Danielewicz1994,DukelskyRoepkeSchuck1998}. Considering symmetric kernels was especially insightful for finding advanced solutions to the quantum many-body problem, with applications ranging from nuclear structure \cite{LitvinovaSchuck2019} to quantum chemistry and condensed matter physics \cite{Tiago2008,Martinez2010,Sangalli2011,SchuckTohyama2016,Olevano2018}, see the recent review \cite{Schuck2021} devoted to a systematic assessment of the EOM method.


Based on the progress of the nuclear many-body theory throughout the decades, recently it was 
demonstrated that beyond-mean-field (BMF) approaches actively employed in nuclear structure physics can be consistently linked to the Hamiltonians operating bare fermionic interactions \cite{LitvinovaSchuck2019,Schuck2019,Schuck2021,LitvinovaSchuck2020,Litvinova2021}.  
Of particular interest are the BMF approaches which account for emergent collective effects of the nuclear medium, first of all, the correlated pairs of quasiparticles known as phonons (vibrations). The phonons are shown to be derived ab-initio, and the phenomenological BMF approaches introducing phonons as part of their input, such as 
the nuclear field theory (NFT) \cite{BohrMottelson1969,BohrMottelson1975,Broglia1976,BortignonBrogliaBesEtAl1977,BertschBortignonBroglia1983}, its variants \cite{Tselyaev1989,LitvinovaTselyaev2007} and the quasiparticle-phonon models (QPM) \cite{Soloviev1992,Malov1976}, follow with the replacement of the bare fermionic interactions by effective interactions derived, e.g., from the density functional theories (DFTs).  The procedure to correct the interaction kernel by subtraction of the Pauli-Villars type \cite{Tselyaev2013} recovering the static limit of the kernel allows one to avoid inconsistencies associated with this replacement in the self-consistent implementations of the response theory \cite{LitvinovaRingTselyaev2007,LitvinovaRingTselyaev2008,LitvinovaRingTselyaev2010,LitvinovaRingTselyaev2013,Tselyaev2018,LitvinovaSchuck2019,Litvinova2023}. 
    
Furthermore, the single-fermion ab-initio EOM method has been formulated for the superfluid case \cite{Litvinova2021a}. Although some versions of such an extension were available in the literature, they were either based on the phenomenological assumptions about the dynamical kernel \cite{VanderSluys1993,Avdeenkov1999,Avdeenkov1999a,BarrancoBrogliaGoriEtAl1999,BarrancoBortignonBrogliaEtAl2005,Tselyaev2007,LitvinovaAfanasjev2011,Litvinova2012,AfanasjevLitvinova2015,IdiniPotelBarrancoEtAl2015} or used truncation on the one-body level to approximate it \cite{Soma2011,Soma2013,Soma2014a,Soma2021}. It was demonstrated in Ref. \cite{Litvinova2021a}, in particular, how pairing correlations beyond the Hartree-Fock-Bogoliubov approximation are integrated into ab-initio theory with the dynamical kernel keeping the two-fermion CFs responsible for the leading effects of emergent collectivity. In contrast to the approaches employing the Bardeen-Cooper-Schrieffer (BCS) or canonical basis, the exact single-fermion EOM was transformed to the basis of the Bogoliubov quasiparticles. This transformation has allowed for consistent unification of the normal and pairing phonon modes and considerable compactification of the superfluid Dyson equation in a more general framework. Remarkably, this enables more efficient handling of the dynamical kernels extending beyond HFB than those of the Gor'kov Green functions and a clear link of the quasiparticle-vibration coupling (qPVC) vertices to the variations of the Bogoliubov quasiparticle Hamiltonian. In Ref. \cite{Litvinova2022} this formulation was employed for the EOM of the response function, which has been worked out in the basis of Bogoliubov quasiparticles from the beginning, leading to a comprehensive ab-initio response theory for superfluid fermionic systems. The qPVC approaches which varying correlation content were derived and shown to generate the known phenomenological approaches, such as the second RPA, NFT, and its extensions, in certain limits. Thus, Ref. \cite{Litvinova2022} has paved the way toward a response theory of spectroscopic accuracy.

In this article, we discuss these recent developments and their implementations for nuclear structure calculations, where the dynamical kernels accounting for emergent collectivity play an important role. In the existing implementations of the nuclear response theory for medium-heavy nuclei, such kernels introduce correlations beyond the simplistic random phase approximation (RPA) and include up to (correlated) two-particle-two-hole ($2p2h$) \cite{LitvinovaRingTselyaev2008,LitvinovaRingTselyaev2010,Gambacurta2011,Gambacurta2015,NiuColoVigezziEtAl2014,NiuNiuColoEtAl2015,RobinLitvinova2016,Tselyaev2018,Robin2019} configuration complexity, in rare cases the $3p3h$ one up to high energy \cite{LitvinovaSchuck2019} or in low-energy limited \cite{Ponomarev1999,LoIudice2012,Savran2011,Tsoneva2019,Lenske:2019ubp} model spaces with the current computational capabilities. These are the approaches mostly based on effective NN-interactions, either schematic or derived from the DFTs, although beyond-RPA approaches employing bare interactions also become available for light and increasingly heavier nuclei \cite{PapakonstantinouRoth2009,Bianco2012,Bacca:2013dma,Knapp:2014xja,Bacca2014,Knapp:2015wpt,DeGregorio2016,DeGregorio2016a,Raimondi2018}. 
An overarching goal is to develop a universal approach demonstrating consistent performance of spectroscopic accuracy across the nuclear chart, and the effort toward it includes advancements in its two major building blocks: (i) the nucleon-nucleon (NN) interactions and (ii) the quantum many-body techniques. In this work, we focus on the latter aspect for accurate modeling of the in-medium dynamics of nucleons using NN interactions as an input to the theory. 


Special emphasis
is put on the recent developments of the nuclear many-body problem elaborated and inspired by the scientific work of Peter Schuck. We derive the exact forms of both the static and dynamical kernels of the one-fermion and two-fermion propagators, from which all the models follow. The major focus is then placed on the 
approximations to the dynamical kernels, where the many-body problem is truncated on the two-body level, i.e., variants of the qPVC kernels. These approximations are found to be the most promising ones for the applications to the regimes of intermediate coupling as they allow for a reasonable compromise between accuracy and feasibility. The latter is possible by making use of modern effective interactions and the former is enabled by the qPVC as the leading approach to emergent collectivity. New numerical implementations for the nuclear response with the self-consistent relativistic qPVC are presented and discussed.

\section{Fermionic propagators in a correlated medium: the two-point functions}
\label{Propagators}
\subsection{Microscopic input and basic definitions}

We adopt a framework, where the many-body fermionic Hamiltonian serves as the only input, which determines uniquely all the properties of the system of interacting fermions. The starting point is the fermionic Hamiltonian in the field-theoretical representation:
\be
H = H^{(1)} + V^{(2)} + W^{(3)} + ... \ .
\label{Hamiltonian}
\ee
The operator $H^{(1)}$ describes the one-body contribution:
\be
H^{(1)} = \sum_{12} t_{12} \psi^{\dag}_1\psi_2 + \sum_{12}v^{(MF)}_{12}\psi^{\dag}_1\psi_2 \equiv \sum_{12}h_{12}\psi^{\dag}_1\psi_2
\label{Hamiltonian1}
\ee
with matrix elements $h_{12}$ combining the kinetic energy $t$ and the mean-field $v^{(MF)}$ part of the interaction. The two-body sector is described by the two-fermion interaction operator $V^{(2)}$:
\be
V^{(2)} = \frac{1}{4}\sum\limits_{1234}{\bar v}_{1234}{\psi^{\dagger}}_1{\psi^{\dagger}}_2\psi_4\psi_3,
\label{Hamiltonian2}
\ee
and $W^{(3)}$ represents the three-body forces, which will be neglected in this work, where we will eventually discuss implementations employing the meson-nucleon interaction in covariant form. In the many-body formalism of this work we will not manifestly use covariant notations, however, keep in mind that the relativistic nucleonic Hamiltonian with the meson-exchange interaction is defined by the terms of essentially the same form as $H^{(1)}$ and $V^{(2)}$ \cite{Brockmann1978,Boyussy1987}. Hence, the general structure of the EOMs remains the same, as it is shown explicitly, for instance, for the one-fermion EOM with the non-symmetric form of the dynamical kernel \cite{Poschenrieder1988,Poschenrieder1988a}. 

The formal covariant theory will be presented elsewhere. Here we utilize the fact that in a relativistic theory, the role of three-body forces is found to be considerably smaller than in a non-relativistic one. The necessity of the relativistic three-body forces in nuclear systems is still debatable, while the corresponding quantitative studies were reported only for few-body systems \cite{Danielewicz1979,Karmanov2011}. We conjecture that the many-body dynamics, non-perturbatively described by the in-medium fermionic propagators in the EOM framework, includes the three- and higher-body forces defined as in Ref. \cite{Karmanov2011} and thus must adequately capture their effects, leaving the contributions associated with the subnucleon degrees of freedom for future studies.
 
The operators $\psi_1$ and $\psi^{\dagger}_1$ in Eqs. (\ref{Hamiltonian1},\ref{Hamiltonian2}) stand for the fermionic fields in some basis of states completely characterized by the number indices. 
In Eq. (\ref{Hamiltonian2}) and in the following we use the antisymmetrized matrix elements ${\bar v}_{1234} = {v}_{1234} - {v}_{1243}$. The fermionic fields obey the anticommutation relations
\bea
[\psi_1,{\psi^{\dagger}}_{1'}]_+ \equiv \psi_1{\psi^{\dagger}}_{1'}  +  {\psi^{\dagger}}_{1'}\psi_1 = \delta_{11'}, \nonumber \\
\left[ \psi_1,{\psi}_{1'} \right]_{+}  = \left[ {\psi^{\dagger}}_1,{\psi^{\dagger}}_{1'}\right]_+ = 0,
\label{anticomm}
\eea
and the Heisenberg form defines their time evolution:
\be
\psi(1) = e^{iHt_1}\psi_1e^{-iHt_1}, \ \ \ \ \ \ {\psi^{\dagger}}(1) = e^{iHt_1}{\psi^{\dagger}}_1e^{-iHt_1}.
\ee

The fermionic in-medium propagator, or real-time Green function, is defined as a correlator of two fermionic field operators:
\be
G(1,1') \equiv G_{11'}(t-t') = -i\langle T \psi(1){\psi^{\dagger}}(1') \rangle,
\label{spgf}
\ee
where $T$ is the chronological ordering operator, and the averaging $\langle ... \rangle$ is performed over the formally exact ground state of the many-body system of $N$ particles. Eq. (\ref{spgf}) describes the propagation of a single fermion in the medium of $N$ interacting fermions.

In the EOM method, it is convenient to use the basis of fermionic states $\{1\}$
which diagonalizes the one-body (single-particle) part of the Hamiltonian $H^{(1)}$: $h_{12} =  \delta_{12}\varepsilon_1$.
The propagator (\ref{spgf}) depends explicitly on a single time difference $\tau = t-t'$, so that its Fourier transform to the energy domain, after inserting the operator $\mathbb{1} = \sum_n |n\rangle\langle n|$ with the complete set of the many-body states, leads to the spectral (Lehmann) representation:
\bea
G_{11'}(\varepsilon) = \sum\limits_{n}\frac{\eta^{n}_{1}\eta^{n\ast}_{1'}}{\varepsilon - (E^{(N+1)}_{n} - E^{(N)}_0)+i\delta} +  \nonumber \\
+ \sum\limits_{m}\frac{\chi^{m}_{1}\chi^{ m\ast}_{1'}}{\varepsilon + (E^{(N-1)}_{m} - E^{(N)}_0)-i\delta}. 
\label{spgfspec}
\eea
$G_{11'}(\varepsilon)$ thus consists of terms of the simple pole character with factorized residues, which is the common feature of the propagators. Its poles are located at the formally exact energies $E^{(N+1)}_{n} - E^{(N)}_0$ and $-(E^{(N-1)}_{m} - E^{(N)}_0)$ of the neighboring $(N+1)$-particle and $(N-1)$-particle systems with respect to the ground state of the background $N$-particle system.
The corresponding residues are the matrix elements of the field operators between the ground state $|0^{(N)}\rangle$ of the reference $N$-particle system and states $|n^{(N+1)} \rangle$ and $|m^{(N-1)} \rangle$:
\be
\eta^{n}_{1} = \langle 0^{(N)}|\psi_1|n^{(N+1)} \rangle , \ \ \ \ \ \ \ \  \chi^{m}_{1} = \langle m^{(N-1)}|\psi_1|0^{(N)} \rangle .
\label{etachi}
\ee
As it follows from their definition, these matrix elements are the weights of the given single-particle (single-hole) configuration on top of the ground state $|0^{(N)}\rangle$ in the $n$-th ($m$-th) state of the $(N+1)$-particle  ($(N-1)$-particle) systems. The residues are thus associated with the occupancies of the corresponding fermionic states.

In the next subsection, we will discuss the EOM for the propagator $G_{11'}(t-t')$ and find that it is connected to the higher-rank two-time (two-point) CFs. Of particular importance are the two two-fermion correlators: the particle-hole propagator, also known as response function, and the particle-particle, or fermionic pair, propagator. 
The response function characterizes the response of the many-body system to an external probe of the one-body character and it is defined as follows:
\bea
R(12,1'2') &\equiv& R_{12,1'2'}(t-t') = -i\langle T\psi^{\dagger}(1)\psi(2)\psi^{\dagger}(2')\psi(1')\rangle \nonumber \\ &=&
-i\langle T(\psi^{\dagger}_1\psi_2)(t)(\psi^{\dagger}_{2'}\psi_{1'})(t')\rangle,
\label{phresp}
\eea
while the pair propagator has the form:
\bea
G(12,1'2') &\equiv& G_{12,1'2'}(t-t') = -i\langle T \psi(1)\psi(2){\psi^{\dagger}}(2'){\psi^{\dagger}}(1')\rangle \nonumber \\ &=& 
-i\langle T(\psi_1\psi_2)(t)(\psi^{\dagger}_{2'}\psi^{\dagger}_{1'})(t')\rangle,
\label{ppGF} 
\eea
where we imply that $t_1 = t_2 = t, t_{1'} = t_{2'} = t'$ and adopt the same phase phase factor as in Eq. (\ref{phresp}) for convenience. 
In analogy to the one-fermion case, inserting the completeness relation between the operator pairs and making the Fourier transformations of these CFs to the energy (frequency) domain lead to:
\be
R_{12,1'2'}(\omega) = \sum\limits_{\nu>0}\Bigl[ \frac{\rho^{\nu}_{21}\rho^{\nu\ast}_{2'1'}}{\omega - \omega_{\nu} + i\delta} -  \frac{\rho^{\nu\ast}_{12}\rho^{\nu}_{1'2'}}{\omega + \omega_{\nu} - i\delta}\Bigr]
\label{respspec}
\ee
\be
G_{12,1'2'}(\omega) =  \sum\limits_{\mu} \frac{\alpha^{\mu}_{21}\alpha^{\mu\ast}_{2'1'}}{\omega - \omega_{\mu}^{(++)}+i\delta} - \sum\limits_{\varkappa} \frac{\beta^{\varkappa\ast}_{12}\beta^{\varkappa}_{1'2'}}{\omega + \omega_{\varkappa}^{(--)}-i\delta}.
\label{resppp}
\ee
Similarly to the one-fermion propagator of Eq. (\ref{spgfspec}), Eqs. (\ref{respspec},\ref{resppp}) satisfy the general requirements of locality and unitarity. The poles  represent the energies $\omega_{\nu} = E_{\nu} - E_0, \omega_{\mu}^{(++)}= E_{\mu}^{(N+2)} - E_0^{(N)}$, and $\omega_{\mu}^{(--)}= E_{\varkappa}^{(N-2)} - E_0^{(N)}$ of the states in the systems with $N$ and $N\pm 2$ particles, respectively. In Eqs. (\ref{spgfspec},\ref{respspec},\ref{resppp})  the sums are formally complete, i.e., run over the discrete spectra and engage the corresponding integrals over the continuum states. 

The  matrix elements in the residues
\bea
\rho^{\nu}_{12} = \langle 0|\psi^{\dagger}_2\psi_1|\nu \rangle \ \ \ \ \ \ \ \ \ \ \ \ \ \ \ \ \ \  \\
\label{trden}
\alpha_{12}^{\mu} = \langle 0^{(N)} | \psi_2\psi_1|\mu^{(N+2)} \rangle  \ \ \ \ \ \ \beta_{12}^{\varkappa} = \langle 0^{(N)} | \psi^{\dagger}_2\psi^{\dagger}_1|\varkappa^{(N-2)} \rangle \nonumber \\
\label{alphabeta}
\eea
are the normal $\rho^{\nu}_{12}$ and anomalous (pairing) $\alpha_{12}^{\mu}, \beta_{12}^{\varkappa}$ transition densities. They give the weights of the pure particle-hole, two-particle and two-hole configurations on top of the ground state $|0^{(N)}\rangle$ in the excited states of the respective systems. These matrix elements play a central role in characterizing transition probabilities, underlying properties of the transitions, pair transfer, and superfluid pairing spectral gaps in nuclear structure applications.

Eqs. (\ref{spgfspec},\ref{respspec},\ref{resppp}) are model-independent, i.e., remain valid for any physical approximations applied to determination of the many-body states $|n\rangle, |m\rangle$, $|\nu\rangle$, $|\mu\rangle$, and $|\varkappa\rangle$. 

As we will see below, the two-fermion two-point functions of Eqs. (\ref{phresp},\ref{ppGF}) will appear in the cluster decomposition of the three-fermion (two-particle-one-hole, or $2p1h$) Green function, which defines the exact symmetric dynamical kernel of the EOM for the one-fermion Green function of Eq. (\ref{spgf}). 

\subsection{Equation of motion for one-fermion propagator }
\label{EOM1}

The EOM for the fermionic propagator (\ref{spgf}) can be generated by taking time derivatives with respect to the times $t$ and $t'$. The detailed derivation procedure is described in Refs. \cite{LitvinovaSchuck2019,Litvinova2021a}, which are in agreement with the major steps given in earlier works \cite{SchuckEthofer1973,Schuck1976,AdachiSchuck1989,Danielewicz1994,DukelskyRoepkeSchuck1998,Dickhoff2004,Dickhoff2005}. 

The differentiation with respect to $t$ leads to
\bea
\partial_t G_{11'}(t-t') = -i\delta(t-t')\langle [\psi_1(t),{\psi^{\dagger}}_{1'}(t')]_+\rangle + \nonumber \\
+ \langle T[H,\psi_1](t){\psi^{\dagger}}_{1'}(t')\rangle, \nonumber\\ 
\label{dtG}                           
\eea
where 
$[H,\psi_1](t) = e^{iHt}[H,\psi_1]e^{-iHt}.$
Evaluating explicitly the commutator with the one-body part of the Hamiltonian and collecting the terms with $G_{11'}(t-t')$, one obtains the equation:
\be
(i\partial_t - \varepsilon_1)G_{11'}(t-t') = \delta_{11'}\delta(t-t') + i\langle T[V,\psi_1](t){\psi^{\dagger}}_{1'}(t')\rangle,
\label{spEOM}
\ee
or, after evaluating the latter commutator, 
\bea
(i\partial_t - \varepsilon_1)G_{11'}(t-t') &=& \delta_{11'}\delta(t-t') \nonumber \\
&+& 
\frac{i}{2}\sum\limits_{ikl}{\bar v}_{i1kl}
\langle T(\psi^{\dagger}_i\psi_l\psi_k)(t){\psi^{\dagger}}_{1'}(t')\rangle \nonumber \\
\label{spEOM1}
\eea
which is commonly referred to as the first EOM, or EOM1. Here we have introduced the Latin dummy indices, which have the same meaning as the number indices, to mark the intermediate fermionic states in the same single-particle basis. The EOM1 of this form can be found in many articles and textbooks on the quantum many-body problem, for instance, in Refs. \cite{SchuckEthofer1973,Dickhoff2005,Poschenrieder1988}. The appearance of the two-fermion CF on the right-hand side of Eq. (\ref{spEOM1}) indicates that the one-fermion propagator and the associated single-particle in-medium trajectories and densities are fundamentally coupled to higher-rank propagators. In principle, an EOM for this two-fermion CF can be generated, but one sees immediately that this EOM further produces a higher-rank CF. The relevance of increasingly-complex CFs to the description of the motion of a single fermion is the characteristic feature of strongly-correlated systems. In weakly-coupled regimes, the perturbation theory is a viable solution discussed in many applications, so we will concentrate on non-perturbative solutions in this work.

Here one can note that the CF in the right-hand side of Eq. (\ref{spEOM1}) 
\be
G^{(2)}_{ilk,1'}(t-t') = -i\langle T(\psi^{\dagger}_i\psi_l\psi_k)(t){\psi^{\dagger}}_{1'}(t')\rangle 
\label{G31}
\ee
depends on the time difference $\tau = t-t'$, so that the Fourier transform of Eq. (\ref{spEOM1}) reads
\be
G_{11'}(\omega) = G^{0}_{11'}(\omega) + \frac{1}{2}\sum\limits_{2ikl}G^{0}_{12}(\omega){\bar v}_{2ikl}G^{(2)}_{ilk,1'}(\omega),
\label{spEOM1a}
\ee
where the free, or uncorrelated, fermionic propagator is introduced as $G^{0}_{11'}(\omega) = \delta_{11'}/(\omega - \varepsilon_1)$.
Eqs. (\ref{spEOM1},\ref{spEOM1a}) can be further transformed to the Dyson form \cite{Dickhoff2004,Dickhoff2005}. There are various possible treatments of the integral part of the Eq. (\ref{spEOM1}), such as the relativistic "$\Lambda^{00},\Lambda^{10},\Lambda^{11} $" approximations \cite{Poschenrieder1988a}, which factorize the CF (\ref{G31}) into two one-fermion CFs, correlated or uncorrelated. Another famous approach is the Gor'kov factorization \cite{Gorkov1958,KucharekRing1991,Litvinova2021a}, which retains, in addition, one-fermion CFs with the same kind of field operators (anomalous Green functions).

More insights into the interacting part of the one-fermion EOM come with its symmetric form, which can be obtained via the second EOM, or EOM2. It is generated by the differentiation of the last term on the right-hand side of Eq. (\ref{spEOM}),
\be
R_{11'}(t-t') =  i\langle T[V,\psi_1](t){\psi^{\dagger}}_{1'}(t')\rangle, 
\ee
with respect to $t'$:
\bea
R_{11'}(t-t')\overleftarrow{\partial_{t'}} 
&=& -i\delta(t-t') \langle \bigl[[V,\psi_1](t),{\psi^{\dagger}}_{1'}(t')\bigr]_+\rangle - \nonumber \\ 
&-& \langle T[V,\psi_1](t)[H,{\psi^{\dagger}}_{1'}](t')\rangle,
\eea
which gives the EOM2:
\bea
R_{11'}(t-t')(-i\overleftarrow{\partial_{t'}}  - \varepsilon_{1'}) &=& -\delta(t-t')\langle \bigl[ [V,\psi_1](t),{\psi^{\dagger}}_{1'}(t')\bigr]_+\rangle\nonumber \\
&+& i\langle T [V,\psi_1](t)[V,{\psi^{\dagger}}_{1'}](t')\rangle.
\label{EOMR}
\eea
Finally, after acting on the EOM1 (\ref{spEOM}) by the operator $(-i\overleftarrow{\partial_{t'}}  - \varepsilon_{1'})$ and Fourier transformation to the energy domain one obtains:
\be
G_{11'}(\omega) 
= G^{0}_{11'}(\omega) + 
\sum\limits_{22'}G^{0}_{12}(\omega)T_{22'}(\omega)G^{0}_{2'1'}(\omega).
\label{spEOM3}
\ee
Thus, the complete in-medium one-fermion propagator is expressed via the free propagator $G^0$ and the $T$-matrix, which absorbs all possible interaction processes of the fermion with the correlated medium. The $T(\omega)$ is the Fourier image of the following time-dependent $T$-matrix:
\bea
T_{11'}(t-t') &=& T^{0}_{11'}(t-t') + T^{r}_{11'}(t-t'), \nonumber\\
T^{0}_{11'}(t-t') &=& -\delta(t-t')\langle \bigl[ [V,\psi_1](t),{\psi^{\dagger}}_{1'}(t')\bigr]_+\rangle, \nonumber \\ 
T^{r}_{11'}(t-t') &=&  i\langle T [V,\psi_1](t)[V,{\psi^{\dagger}}_{1'}](t')\rangle.
\label{Toperator}
\eea
The superscript "0" marks the static, or instantaneous, part of the $T$-matrix, and "r" is associated with its dynamical, or time-dependent, part containing retardation effects. The EOM (\ref{spEOM3}) in the operator form reads:
\be
G(\omega) = G^{0}(\omega) + G^{0}(\omega)T(\omega)G^{0}(\omega).
\label{spEOM4}
\ee
It is often more instructive to have this EOM in the Dyson ansatz, for which one introduces the irreducible part of the $T$-matrix with respect to the uncorrelated one-fermion propagator $G^{0}$, the self-energy (called also interaction kernel): $\Sigma = T^{irr}$. This operation removes all the contributions containing parts connected by the 
one-fermion free propagator $G^0$ from the $T$-matrix by the following definition:
\be
T(\omega) = \Sigma(\omega) + \Sigma(\omega) G^{0}(\omega)T(\omega).
\label{DysonT}
\ee
Combining Eqs. (\ref{spEOM4}) and (\ref{DysonT}), one arrives at the Dyson equation for the fermionic propagator:
\be
G(\omega) = G^{0}(\omega) + G^{0}(\omega)\Sigma(\omega) G(\omega).
\label{Dyson}
\ee
The self-energy holds the same decomposition as the $T$-matrix:
\be
\Sigma_{11'}(\omega) = \Sigma_{11'}^{0} + \Sigma_{11'}^{r}(\omega),
\label{Somega}
\ee
where
\be
 \Sigma^{0}_{11'} = -\langle[[V,\psi_1],{\psi^{\dagger}}_{1'}]_+\rangle = \sum\limits_{il}{\bar v}_{1i1'l}\rho_{li}, 
 \label{MF}
\ee
with  $\rho_{li} = \langle{\psi^{\dagger}}_i\psi_l\rangle$  being the ground-state one-body density, and $\Sigma_{11'}^{r}(\omega)$ is the Fourier image of
\bea
\Sigma_{11'}^{r}(t-t') &=& -\frac{i}{4} \sum\limits_{npq}\sum\limits_{ikl}{\bar v}_{1ikl}\times \nonumber\\
&\times&\langle T \bigl(\psi^{\dagger}_i\psi_l\psi_k\bigr)(t)\bigl(\psi^{\dagger}_p\psi^{\dagger}_q\psi_n\bigr)(t')\rangle^{irr}
{\bar v}_{qpn1'}  \nonumber\\
&=& \frac{1}{4} \sum\limits_{npq}\sum\limits_{ikl}{\bar v}_{1ikl}
G^{(pph)irr}_{ilk,nqp}(t-t'){\bar v}_{qpn1'}.\nonumber\\
\label{Tr}
\eea


Thus, the EOM (\ref{Dyson}) for the fermionic propagator $G(\omega)$  is formally a closed equation with respect to $G(\omega)$. However, its self-energy, which in this version has the symmetric form of a CF "sandwiched" between two interaction matrix elements, contains the three-fermion Green function. The obtained form of the self-energy (\ref{Somega}-\ref{Tr}) indicates a clear separation between the static (instantaneous) Hartree-Fock-like term (\ref{MF}) and the dynamical term (\ref{Tr}), which accumulates all the retardation effects. 
Accordingly, the former is responsible for the short-range and the latter generates long-range correlations. Here the short range is associated with the range of the input bare interaction $\bar v$, and the long range may extend to the size of the entire many-body system.

So far the theory is exact, but again, the EOM for the three-body propagator generates higher-rank propagators, which makes the exact solution of the many-body problem hardly tractable. There are various ways to approximate the three-fermion propagator in the dynamical kernel of the one-fermion EOM. Some approximations can be obtained by making use of the cluster decomposition of the CF in the dynamical kernel (\ref{Tr}). In a symbolic form, it reads:
\be
G^{(pph)irr} = G^{(p)}G^{(p)}G^{(h)} + G^{(p)}R^{(ph)} + G^{(h)}G^{(pp)} + \sigma^{(pph)},
\label{CD}
\ee
where the number of particles (holes) in the superscripts indicates the rank of the respective CF and the sum implicitly includes all the necessary antisymmetrizations, see Refs. \cite{VinhMau1969,Mau1976,LitvinovaSchuck2019} for details. Retaining the first term only is the approach, which truncates the many-body problem at the one-body level, which is sometimes called the self-consistent Green functions approach. Some implementations were presented in Refs. \cite{Poschenrieder1988,Poschenrieder1988a}. The one-fermion EOM has then a closed form and can, in principle, be solved iteratively. The decomposition retaining all possible terms with one-fermion and two-fermion propagators was discussed in detail, for instance, in Refs. \cite{Martin1959,VinhMau1969,Mau1976,RingSchuck1980,Rijsdijk1992,LitvinovaSchuck2019}. It was shown, in particular, that this approximation can be linked to the nuclear field theory \cite{BohrMottelson1969,BohrMottelson1975,Broglia1976,BortignonBrogliaBesEtAl1977,BertschBortignonBroglia1983}, which implies a coupling between particles and phonons of both particle-hole and particle-particle origins. In this approximation, the one-fermion dynamical kernel takes the form
\begin{eqnarray}
  \Sigma^{r}_{11'}(\omega) =  \Sigma^{r(ph)}_{11'}(\omega) +  \Sigma^{r(pp)}_{11'}(\omega) +  \Sigma^{r(0)}_{11'}(\omega),\nonumber \\
 \label{SEirr2}
 \end{eqnarray}
where 
\be
\Sigma^{r(ph)}_{11'}(\omega) = -\sum\limits_{33'}\int\limits_{-\infty}^{\infty}\frac{d\varepsilon}{2\pi i} \Gamma^{ph}_{13',1'3}(\omega - \varepsilon)G_{33'}(\varepsilon),
\label{FISphe}
\ee
\be
\Sigma^{r(pp)}_{11'}(\omega) = \sum\limits_{22'}\int\limits_{-\infty}^{\infty}\frac{d\varepsilon}{2\pi i} \Gamma^{pp}_{12,1'2'}(\omega + \varepsilon)G_{2'2}(\varepsilon),
\ee
\bea
\Sigma^{r(0)}_{11'}(\omega) &=& -\sum\limits_{2342'3'4'}{\bar v}_{1234}\nonumber\\
&\times~&\int\limits_{-\infty}^{\infty}\frac{d\varepsilon d\varepsilon'}{(2\pi i)^2}  G_{44'}(\omega+\varepsilon'-\varepsilon)G_{33'}(\varepsilon)G_{2'2}(\varepsilon')
\nonumber \\
&\times&{\bar v}_{4'3'2'1'},
\label{SEdyn0}
\eea
and the amplitudes $\Gamma^{ph}$ and $\Gamma^{pp}$ are defined as
\bea
\Gamma^{ph}_{13',1'3} =  \sum\limits_{242'4'}{\bar v}_{1234}R^{(ph)}_{24,2'4'}(\omega){\bar v}_{4'3'2'1'} = \nonumber \\ = 
\sum\limits_{\nu,\sigma=\pm1} g^{{\nu}(\sigma)}_{13}D^{(\sigma)}_{\nu}(\omega)g^{\nu(\sigma)\ast}_{1'3'},
\label{mappingph}
\eea
where we introduced the phonon vertices $g^{\nu}$ and propagators $D_{\nu}(\omega)$:
\bea
g^{\nu(\sigma)}_{13} = \delta_{\sigma,+1}g^{\nu}_{13} + \delta_{\sigma,-1}g^{\nu\ast}_{31}, \ \ \ \ 
g^{\nu}_{13} = \sum\limits_{24}{\bar v}_{1234}\rho^{\nu}_{42}, 
\label{vert_ph} \nonumber\\
\\
D_{\nu}^{(\sigma)}(\omega) = \frac{\sigma}{\omega - \sigma(\omega_{\nu} - i\delta)}, \ \ \ \
\omega_{\nu} = E_{\nu} - E_0, \nonumber \\
\label{gDPVCph}
\eea 
and
\bea
\Gamma^{pp}_{12,1'2'}(\omega) = \sum\limits_{343'4'}{v}_{1234}G^{(pp)}_{43,3'4'}(\omega){v}_{4'3'2'1'} = \nonumber \\
= \sum\limits_{\mu,\sigma=\pm1} \gamma^{\mu(\sigma)}_{12}\Delta^{(\sigma)}_{\mu}(\omega)\gamma^{\mu(\sigma)\ast}_{1'2'}
\label{mappingpp}
\eea
with the pairing vertices $\gamma^{\mu(\pm)}$ and propagator $\Delta_{\mu}(\omega)$
\be
\gamma^{\mu(+)}_{12} = \sum\limits_{34} v_{1234}\alpha_{34}^{\mu}, \ \ \ \ \ \ \gamma_{12}^{\varkappa(-)} = \sum\limits_{34}\beta_{34}^{\varkappa}v_{3412}. 
\label{vert_pp}
\ee
\be
\Delta^{(\sigma)}_{\mu}(\omega) = \frac{\sigma}{\omega - \sigma(\omega_{\mu}^{(\sigma\sigma)} - i\delta)}.
\ee

\begin{figure}
\begin{center}
\includegraphics[scale=0.52]{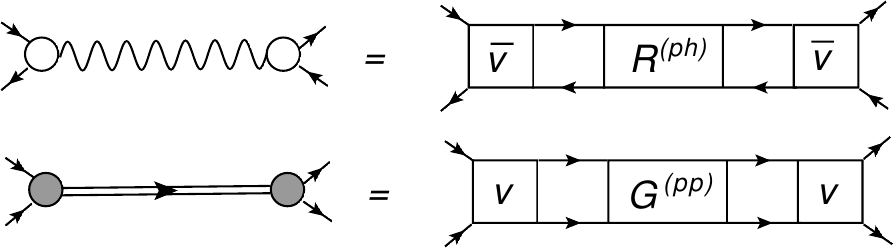}
\end{center}
\caption{The microscopic mechanism of the quasiparticle-vibration coupling:  the phonon vertices are denoted by the empty and filled circles, their propagators correspond to the wavy and double lines, for the normal (top) and pairing, or superfluid (bottom), phonons, respectively. The bare interaction is given by the squares, antisymmetrized $\bar v$ and plain $v$, and the two-fermion correlation functions are the rectangular blocks $R^{(ph)}$ and $G^{(pp)}$. Solid lines with arrows are associated with fermionic particle (right) and hole (left) states with respect to the many-body $N$-particle ground state $|0^{(N)}\rangle$. The figure is adopted from Ref. \cite{LitvinovaSchuck2019}.}
\label{PVCmap}%
\end{figure}

In the definitions above, the particle-particle ($pp$) and particle-hole ($ph$) correlation functions defined by Eqs. (\ref{phresp},\ref{ppGF}) are employed, while the mappings of Eqs. (\ref{mappingph},\ref{mappingpp}) to the particle-vibration coupling (PVC) are displayed diagrammatically in Fig. \ref{PVCmap}.
\begin{figure*}
\begin{center}
\includegraphics*[scale=0.75]{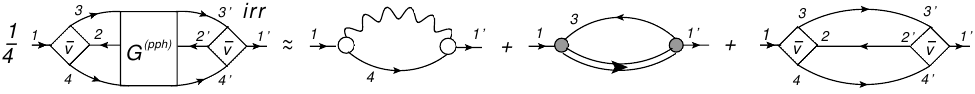}
\end{center}
\caption{The dynamical kernel $\Sigma^{r}$ of Eq. (\ref{SEirr2}) in terms of the particle-vibration coupling, with the same conventions as in Fig. \ref{PVCmap}. The block $G^{(pph)}$ stands for the three-fermion propagator of Eq. (\ref{Tr}). The figure is adopted from Ref. \cite{LitvinovaSchuck2019}.}
\label{SEdyn}%
\end{figure*}
Thus,  Eq. (\ref{SEirr2}) is the foundation for microscopic approaches to the single-particle self-energy, which refer to the phenomenon of PVC. The mappings (\ref{mappingph},\ref{mappingpp}) lead to the diagrammatic form of the self-energy shown in Fig. \ref{SEdyn}. The spectral representations of the respective terms read:
\bea
\Sigma^{r(ph)}_{11'}(\omega) = \sum\limits_{33'} \Bigl[ 
\sum\limits_{\nu n}\frac{\eta_3^{n}{g}_{13}^{\nu}{g}_{1'3'}^{\nu\ast}\eta_{3'}^{n\ast}}{\omega - \omega_{\nu} - \varepsilon_n^{(+)} + i\delta} +
\nonumber\\ +
\sum\limits_{\nu m} \frac{\chi_3^{m} g_{31}^{\nu\ast}g_{3'1'}^{\nu}\chi_{3'}^{m\ast}}{\omega + \omega_{\nu} + \varepsilon_m^{(-)} - i\delta} 
\Bigr],
\label{Sigmarph}
\eea
\bea
\Sigma^{r(pp)}_{11'}(\omega) = \sum\limits_{22'} \Bigl[ \sum\limits_{\mu m} \frac{\chi_2^{m\ast} \gamma_{12}^{\mu(+)}\gamma_{1'2'}^{\mu(+)\ast}\chi_{2'}^{m}}{\omega - \omega_{\mu}^{(++)} - \varepsilon_m^{(-)} + i\delta} + \nonumber\\
+ \sum\limits_{\varkappa n}\frac{\eta_2^{n\ast}{\gamma}_{21}^{\varkappa(-)\ast}{\gamma}_{2'1'}^{\varkappa(-)}\eta_{2'}^n}{\omega + \omega_{\varkappa}^{(--)} + \varepsilon_n^{(+)} - i\delta} \Bigr],
\label{FISrpp}
\eea
%
\bea
&\Sigma&^{r(0)}_{11'}(\omega) = -\sum\limits_{2342'3'4'} {\bar v}_{1234}\times \nonumber \\ &\times&\Bigl[\sum\limits_{mn'n''} \frac{\chi_{2'}^{m}\chi_2^{m\ast}\eta_3^{n'}\eta_{3'}^{n'\ast}\eta_4^{n''}\eta_{4'}^{n''\ast}}
{\omega - \varepsilon_{n'}^{(+)} - \varepsilon_{n''}^{(+)} - \varepsilon_{m}^{(-)} + i\delta} \nonumber \\
&+& \sum\limits_{nm'm''} \frac{\eta_{2'}^{n}\eta_{2}^{n\ast}\chi_3^{m'}\chi_{3'}^{m'\ast}\chi_4^{m''}\chi_{4'}^{m''\ast}}
{\omega + \varepsilon_{n}^{(+)} + \varepsilon_{m'}^{(-)} + \varepsilon_{m''}^{(-)} - i\delta} \Bigr] {\bar v}_{4'3'2'1'}. 
\label{Sigmar0}
\eea
Here the single-particle energies in the neighboring $(N+1)$-particle system are defined as $\varepsilon_n^{(+)} = E^{(N+1)}_{n} - E^{(N)}_0$ and those in the $(N-1)$-particle system as $\varepsilon_m^{(-)} = E^{(N-1)}_{m} - E^{(N)}_0$. 

The complete dynamical part  (\ref{SEirr2}-\ref{SEdyn0}) of the fermionic self-energy truncated at the two-body level is shown in Fig. \ref{SEdyn} in the diagrammatic form.   Note that the signs in front of the diagrams are convention dependent, for instance, they may not respect Feynman's convention. For instance, the last uncorrelated term is often shown with the "-" sign in the literature, and the phonon vertices may include the multiplier "$i$". 
The first two diagrams on the right-hand side in Fig. \ref{SEdyn} are the typical one-loop diagrams, which are analogous to the electron self-energy correction in quantum electrodynamics (QED). In electronic systems, an electron typically emits and reabsorbs a virtual photon or a phonon excitation of the lattice in a metal. In the nucleonic self-energy of quantum hadrodynamics (QHD) a single nucleon emits and reabsorbs mesons of various kinds, which in the lowest order can be described by the first diagram on the right-hand side if the wavy line is attributed to the meson propagator. In the applications discussed in this work, the meson exchange acts as a bare interaction (although slightly renormalized) described by the matrix elements of $\bar v$.  The dynamical self-energy then expresses the medium-induced emerging contribution to the fermionic interaction. Quite remarkably, in the leading approximation dominated by the term $\Sigma^{r(ph)}$ (\ref{Sigmarph}) in a strong coupling regime, the dynamical fermionic self-energy takes an analogous form of the boson exchange, thereby illustrating how this type of interaction is driven across the energy scales. This process can be expressed by an effective Hamiltonian with the explicit phonon degrees of freedom, as it is done in the phenomenological NFT. The difference between the nuclear system and QED or QHD, besides the differences in the reference ("vacuum") states, is that in the latter theories, the fermionic and bosonic degrees of freedom are independent. Interestingly, the PVC vertices in nuclear systems are not the effective parameters of the theory but are in principle calculable from the underlying fermionic bare interaction. 


There are very few realizations of the PVC model based on the bare NN interaction \cite{Barbieri2009,Barbieri2009a}. The implementations employing 
effective interactions \cite{LitvinovaRing2006,Litvinova2012,LitvinovaAfanasjev2011,AfanasjevLitvinova2015} are more accurate quantitatively and more feasible because the phonons can be reasonably described already on the level of (quasiparticle) random phase approximation ((Q)RPA) and successive iterations of the fermionic propagator in the Dyson equation may be omitted.  However, such approaches inevitably imply an additional procedure to remove the double counting of PVC, implicitly contained in the effective interaction \cite{Tselyaev2013}. An elegant way of avoiding such double counting is the explicit subtraction of the dynamical PVC kernel taken in the static limit from the effective interaction.  The subtraction method is widely applied in calculations of two-nucleon Green functions, in particular, the particle-hole response \cite{Tselyaev2013,LitvinovaTselyaev2007,LitvinovaRingTselyaev2007,LitvinovaRingTselyaev2008,Gambacurta2015,LitvinovaSchuck2019}, while a corresponding method has not been yet adopted to the case of the one-body propagator.

The information about the two-fermion propagators $R^{(ph)}$ and $G^{(pp)}$ in terms of the phonon vertices and propagators can be obtained from their direct computation by solving the EOMs for these correlation functions. The theory and implementations of the corresponding EOMs are presented in Refs. \cite{DukelskyRoepkeSchuck1998,Olevano2018,LitvinovaSchuck2019,LitvinovaSchuck2020}.


The spectral forms (\ref{Sigmarph}-\ref{Sigmar0}) of the three terms with the explicit locality and unitarity are best suited for the accurate diagrammatic mapping and they reveal a different sign of the "second-order" term $\Sigma^{r(0)}$ as compared to the "radiative correction" terms $\Sigma^{r(ph)}$ and $\Sigma^{r(pp)}$ containing phonons. This means, in particular, that potentially the positivity can be violated in the optical theorem. As it was pointed out, for instance, in Refs. \cite{Schuck1973,Danielewicz1994}, to prevent this violation it is important to keep the integrity of the spectral representation of the dynamical self-energy (\ref{Tr})
\bea
\Sigma_{11'}^{r}(\omega) &=& \frac{1}{4} \sum\limits_{rpq}\sum\limits_{ikl}{\bar v}_{1ikl}
G^{(pph)irr}_{ilk,rqp}(\omega){\bar v}_{qpr1'}\nonumber\\
\label{Tromega}
G^{(pph)}_{ilk,rqp} (\omega) &=&  \sum\limits_{n}\frac{\langle 0|\psi^{\dagger}_i\psi_{l}\psi_{k}|n\rangle\langle n|\psi^{\dagger}_{p}\psi^{\dagger}_{q}\psi_{r}|0\rangle}{\omega - (E^{(N+1)}_{n} - E^{(N)}_0)+i\delta} \nonumber \\
&+& \sum\limits_{m}\frac{\langle 0| \psi^{\dagger}_{p}\psi^{\dagger}_{q}\psi_{r}|m\rangle\langle m|\psi^{\dagger}_i\psi_{l}\psi_{k} |0\rangle}{\omega + (E^{(N-1)}_{m} - E^{(N)}_0)-i\delta}. \nonumber \\ 
\label{pphgfspec}
\eea
Comparing the exact dynamical self-energy of Eq. (\ref{pphgfspec}) to Eqs. (\ref{Sigmarph}-\ref{Sigmar0}), one can see clearly that the above-mentioned sign problem and potentially other inconsistencies may appear because of the different analytical structures of the exact and approximate solutions. 

One may notice that the poles of the $G^{(pph)}$ in Eq. (\ref{pphgfspec}) coincide with those of the one-fermion propagator $G$ in the form of Eq. (\ref{spgfspec}), because both sums on the right-hand sides of these expressions run over the complete formally exact spectra of the same $N\pm 1$ systems. Combining these equations with Eq. (\ref{spEOM3}) and considering the energy argument close to the exact pole $\varepsilon_n$, one obtains:
\be
\langle 0|\psi_1|n\rangle = \frac{1}{2}\sum\limits_{ikl}{\bar v}_{1ikl}\frac{\langle 0|\psi^{\dagger}_i\psi_l\psi_k|n\rangle}{\varepsilon_n - \varepsilon_1},
\ee
which is consistent with the exact EOM for a single field operator \cite{Zelevinsky2017}.

Considerable effort on finding viable approximations for the irreducible part of $G^{(pph)}$ compatible with the spectral expansion (\ref{pphgfspec}) was undertaken in the past. The authors of Ref. \cite{Schuck1973} have formulated the two-particle-one-hole ($2p1h$) RPA for this correlation function, while a more accurate approximation to the $2p1h$ energies and matrix elements via Faddeev series has been developed in Ref. \cite{Danielewicz1994} to include emergent  collectivity in 
the $ph$ and $pp$ channels.
 Nuclear structure implementations, however, are still quite limited and confined by the Tamm-Dancoff and random phase approximations to the $ph$ and $pp$ correlation functions \cite{Rijsdijk1992,Rijsdijk1996,Barbieri2009,Barbieri2009a}. 

\subsection{Superfluid phase}
\label{superfluid}

The majority of strongly-correlated fermionic systems, including atomic nuclei, exhibit pronounced effects of superfluidity \cite{RingSchuck1980}, that need an extended treatment. The superfluid phase is generally characterized by the enhanced formation of Cooper pairs and pairing phonons, which appear to be a dynamical counterpart of the Cooper pairs. In calculations for normal systems within the PVC approach to the self-energy using effective interactions one usually neglects the pairing phonons because of their relatively low importance, but they should be kept for superfluid systems and also if a bare interaction is used. 

Within the PVC approach discussed above the pairing interaction is fully dynamical and mediated by the pairing phonons emerging naturally in the one-fermion self-energy. In the traditional frameworks based on effective interactions, however, the pairing is included in the static approximations, such as the BCS or the Hartree-(Fock)-Bogoliubov (HFB) ones. On this level of description, the corresponding Green function technique is the Gor'kov Green functions, which extends the notion of the one-fermion propagator (\ref{spgf}) by introducing anomalous propagators with the same kind of fermionic operators, see Eq. (\ref{GG}) below. These correlation functions do not vanish because correlated fermionic pairs are present in the ground state of the system.

The simplest approach to the Gor'kov Green functions can be obtained from the EOM1 if the two-body correlations are neglected, see, for instance, \cite{Gorkov1958}. In ref. \cite{Litvinova2021a} it is shown how the Gor'kov theory is generalized beyond this approximation, in particular, to the inclusion of the PVC effects in the dynamical self-energy.
It is convenient to introduce the HFB basis, or the basis of the Bogoliubov quasiparticles \cite{Bogolubov1947}. The states in this basis 
combine particle and hole states, i.e., the states above and below the Fermi energy: 
\bea
\psi_1 = U_{1\mu}\alpha_{\mu} + V^{\ast}_{1\mu}\alpha^{\dagger}_{\mu}\nonumber\\
\psi^{\dagger}_1 = V_{1\mu}\alpha_{\mu} + U^{\ast}_{1\mu}\alpha^{\dagger}_{\mu}.
\label{Btrans}
\eea
Here and henceforth the Greek indices are used to denote fermionic states in the HFB basis, while the number indices and the Roman indices are reserved for the single-particle mean-field basis states. The repeated indices $\mu$ imply summation, so that Eq. (\ref{Btrans}) can be expressed in a matrix form:
\bea
\left( \begin{array}{c} \psi \\ \psi^{\dagger} \end{array} \right) = \cal{W} \left( \begin{array}{c} \alpha \\ \alpha^{\dagger} \end{array} \right),
\eea
where
\bea
\cal{W} = \left( \begin{array}{cc} U & V^{\ast} \\ V & U^{\ast} \end{array} \right) \ \ \ \ \ \  \cal{W}^{\dagger} = \left( \begin{array}{cc} U^{\dagger} & V^{\dagger} \\ V^T & U^T \end{array} \right) 
\eea
are unitary matrices. The quasiparticle operators $\alpha$ and $\alpha^{\dagger}$ obey the same anticommutator algebra as the particle operators $\psi$ and $\psi^{\dagger}$, while the matrices $U$ and $V$ satisfy:
\bea
U^{\dagger}U + V^{\dagger}V = \mathbb{1}\ \ \ \ \ \ UU^{\dagger} + V^{\ast}V^{T} = \mathbb{1}\nonumber\\
U^TV + V^TU = 0\ \ \ \ \ \  UV^{\dagger} + V^{\ast}U^{T} = 0 .
\label{UV}
\eea

The generalized fermionic propagator, therefore, takes the form
\bea
{\hat G}_{12}(t-t') = -i\langle T\Psi_1(t)\Psi^{\dagger}_2(t')\rangle = \nonumber\\
= -i\theta(t-t')\left( \begin{array}{cc} \langle \psi_1(t)\psi^{\dagger}_2(t')\rangle &  \langle \psi_1(t)\psi_2(t')\rangle \\
 \langle \psi^{\dagger}_1(t)\psi^{\dagger}_2(t')\rangle &  \langle \psi^{\dagger}_1(t)\psi_2(t')\rangle
\end{array} \right)  + \nonumber \\
+ i\theta(t'-t)\left( \begin{array}{cc} \langle \psi^{\dagger}_2(t')\psi_1(t)\rangle &  \langle \psi_2(t')\psi_1(t)\rangle \\
 \langle \psi^{\dagger}_2(t')\psi^{\dagger}_1(t)\rangle &  \langle \psi_2(t')\psi^{\dagger}_1(t)\rangle
\end{array} \right) \nonumber\\
= \left( \begin{array}{cc} G_{12}(t-t')  &  F^{(1)}_{12}(t-t') \\ 
F^{(2)}_{12}(t-t') & G^{(h)}_{12}(t-t')\end{array}\right). 
\label{GG}
\eea
with
\be
\Psi_1(t_1) = \left( \begin{array}{c} \psi_1(t_1) \\ \psi_1^{\dagger}(t_1) \end{array} \right), \ \ \ \ \ \ \ \ \ 
\Psi^{\dagger}_1(t_1) = \Bigl( \psi_1^{\dagger}(t_1) \ \ \ \ \ \psi_1(t_1) \Bigr).
\label{psicol}
\ee
In Ref. \cite{Litvinova2021a} it was shown in detail how the procedure of generating the one-fermion EOM is performed for all the components of the propagator (\ref{GG}). A unified EOM was obtained, in particular, by transforming the matrix equation for the ${\hat G}_{12}$ to the quasiparticle basis. 
The resulting Gor'kov-Dyson equation for the quasiparticle propagator in the energy domain reads:
\be
G^{(\eta)}_{\nu\nu'}(\varepsilon) = {\tilde G}^{(\eta)}_{\nu\nu'}(\varepsilon) + \sum\limits_{\mu\mu'}{\tilde G}^{(\eta)}_{\nu\mu}(\varepsilon)\Sigma^{r(\eta)}_{\mu\mu'}(\varepsilon)G^{(\eta)}_{\mu'\nu'}(\varepsilon). 
\label{Dyson_qp}
\ee
The indices $\eta = +$ and $\eta = -$ are introduced for the quasiparticle forward and backward components, respectively. The mean-field ${\tilde G}^{(\eta)}_{\nu\nu'}(\varepsilon)$ and exact $G^{(\eta)}_{\nu\nu'}(\varepsilon)$ propagator components are defined as follows:
\be
{\tilde G}^{(\eta)}_{\nu\nu'}(\varepsilon) = \frac{\delta_{\nu\nu'}}{\varepsilon - \eta(E_{\nu} - E_0 - i\delta)}
\label{MFG_qp}
\ee
\be
{G}^{(\eta)}_{\nu\nu'}(\varepsilon) = \sum\limits_n\frac{S^{\eta(n)}_{\nu\nu'}}{\varepsilon - \eta(E_{n} - E_0 - i\delta)},
\label{G_qp}
\ee
with the mean-field energies of the Bogoliubov quasiparticles $E_{\nu}$ and formally exact quasiparticle energies $E_{n}$.
The residues are the matrix element products $S^{+(n)}_{\nu\nu'} = \langle 0|\alpha_{\nu}|n\rangle\langle n|\alpha^{\dagger}_{\nu'}|0\rangle$ and $S^{-(m)}_{\nu\nu'} = \langle 0|\alpha_{\nu}|m\rangle\langle m|\alpha^{\dagger}_{\nu'}|0\rangle$ with the formally exact states $|n\rangle$ and $|m\rangle$. 
The components of the dynamical kernel in the quasiparticle basis are related to those in the single-particle basis as follows:
\bea
\Sigma^{r(+)}_{\mu\mu'}(\varepsilon) &=& \sum\limits_{12} \Bigl(U^{\dagger}_{\mu 1} \ \ \  V^{\dagger}_{\mu 1} \Bigr) \left( \begin{array}{cc} \Sigma^r_{12}(\varepsilon) &  \Sigma^{(1)r}_{12}(\varepsilon)\\ \Sigma^{(2)r}_{12}(\varepsilon) & \Sigma^{(h)r}_{12}(\varepsilon)\end{array}\right)
\left( \begin{array}{c} U_{2\mu'} \\ V_{2\mu'} \end{array}\right)  \nonumber\\
&=& \sum\limits_{12} \Bigl( U^{\dagger}_{\mu 1}\Sigma^r_{12}(\varepsilon)U_{2\mu'} + U^{\dagger}_{\mu 1}\Sigma^{(1)r}_{12}(\varepsilon)V_{2\mu'} \nonumber\\
&+& V^{\dagger}_{\mu 1}\Sigma^{(2)r}_{12}(\varepsilon)U_{2\mu'} + V^{\dagger}_{\mu 1}\Sigma^{(h)r}_{12}(\varepsilon)V_{2\mu'} \Bigr),
\label{Sigma+}
\eea
\bea
\Sigma^{r(-)}_{\mu\mu'}(\varepsilon) &=& \sum\limits_{12} \Bigl(V^{T}_{\mu 1} \ \ \  U^{T}_{\mu 1} \Bigr) \left( \begin{array}{cc} \Sigma^r_{12}(\varepsilon) &  \Sigma^{(1)r}_{12}(\varepsilon)\\ \Sigma^{(2)r}_{12}(\varepsilon) & \Sigma^{(h)r}_{12}(\varepsilon)\end{array}\right)
\left( \begin{array}{c} V^{\ast}_{2\mu'} \\ U^{\ast}_{2\mu'} \end{array}\right)  \nonumber\\
&=& \sum\limits_{12} \Bigl( V^{T}_{\mu 1}\Sigma^r_{12}(\varepsilon)V^{\ast}_{2\mu'} + V^{T}_{\mu 1}\Sigma^{(1)r}_{12}(\varepsilon)U^{\ast}_{2\mu'} \nonumber\\
&+& U^{T}_{\mu 1}\Sigma^{(2)r}_{12}(\varepsilon)V^{\ast}_{2\mu'} + U^{T}_{\mu 1}\Sigma^{(h)r}_{12}(\varepsilon)U^{\ast}_{2\mu'} \Bigr),
\label{Sigma-}
\eea
while the matrix structure of the dynamical self-energy ${\hat\Sigma}_{11'}^{r}$ corresponds to the structure of the propagator matrix (\ref{GG}): 
\be
{\hat\Sigma}_{12}^{r} (\varepsilon) = \left( \begin{array}{cc} \Sigma^r_{12}(\varepsilon) &  \Sigma^{(1)r}_{12}(\varepsilon)\\ \Sigma^{(2)r}_{12}(\varepsilon) & \Sigma^{(h)r}_{12}(\varepsilon)\end{array}\right).
\ee
\begin{figure*}
\begin{center}
\vspace{0.3cm}
\includegraphics[scale=0.8]{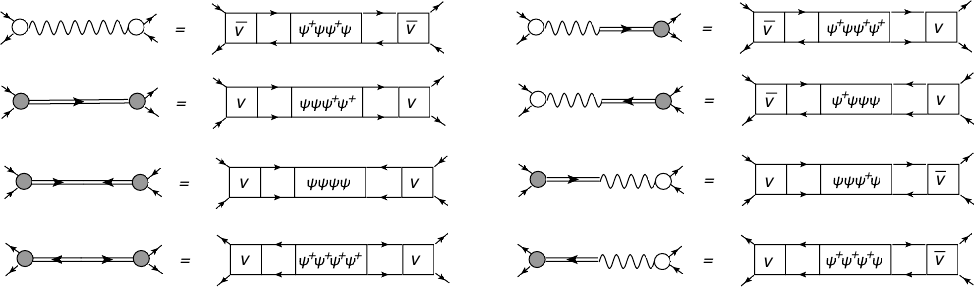}
\end{center}
\caption{The emergence of the quasiparticle-vibration coupling amplitudes in the diagrammatic form. Same conventions as in Figs. \ref{PVCmap}, \ref{SEdyn} apply to the normal and pairing vibration (phonon) vertices and propagators, and for the antisymmetrized $\bar v$ and non-antisymmetrized $v$ interaction matrix elements. The operator products in the rectangular boxes, together with the attached fermionic lines (solid lines with arrows), denote the two-point correlation functions, according to the rule: \framebox{abcd} $= -i\langle T(ab)(t)(cd)(t')\rangle$. The figure is adopted from Ref.  \cite{Zhang2022}.}
\label{QVC_map}
\end{figure*}

The components of the exact dynamical self-energy are obtainable as the Fourier images of the double contractions of the three-fermion correlators with two interaction matrix elements:
\bea
\Sigma_{11'}^{r}(t-t') &=& \frac{i}{4} \sum\limits_{ikl}\sum\limits_{mnq}{\bar v}_{1ikl} \nonumber\\
&\times&\langle T \bigl(\psi^{\dagger}_i\psi_l\psi_k\bigr)(t)\bigl(\psi^{\dagger}_m\psi^{\dagger}_n\psi_q\bigr)(t')\rangle^{irr}
{\bar v}_{mnq1'} \nonumber\\
\Sigma_{11'}^{(1)r}(t-t') 
&=& \frac{i}{4}\sum\limits_{ikl}\sum\limits_{mnq}{\bar v}_{1ikl} \nonumber\\
&\times&\langle T(\psi^{\dagger}_i\psi_l\psi_k )(t)(\psi^{\dagger}_m\psi_q\psi_n)(t')\rangle^{irr} {\bar v}_{1'mnq}. \nonumber \\
\Sigma_{11'}^{(2)r}(t-t') 
&=& \frac{i}{4}\sum\limits_{ikl}\sum\limits_{mnq}{\bar v}_{ikl1} \nonumber\\
&\times&\langle T(\psi^{\dagger}_i\psi^{\dagger}_k\psi_l )(t)(\psi^{\dagger}_m\psi^{\dagger}_n\psi_q)(t')\rangle^{irr} {\bar v}_{mnq1'}\nonumber \\
\Sigma_{11'}^{(h)r}(t-t') 
&=& \frac{i}{4}\sum\limits_{ikl}\sum\limits_{mnq}{\bar v}_{ikl1} \nonumber\\
&\times&\langle T(\psi^{\dagger}_i\psi^{\dagger}_k\psi_l )(t)(\psi^{\dagger}_m\psi_q\psi_n)(t')\rangle^{irr} {\bar v}_{1'mnq}. \nonumber \\
\label{Tsfdyn}
\nonumber\\
\eea
%
\begin{figure*}
\begin{center}
\includegraphics[scale=0.34]{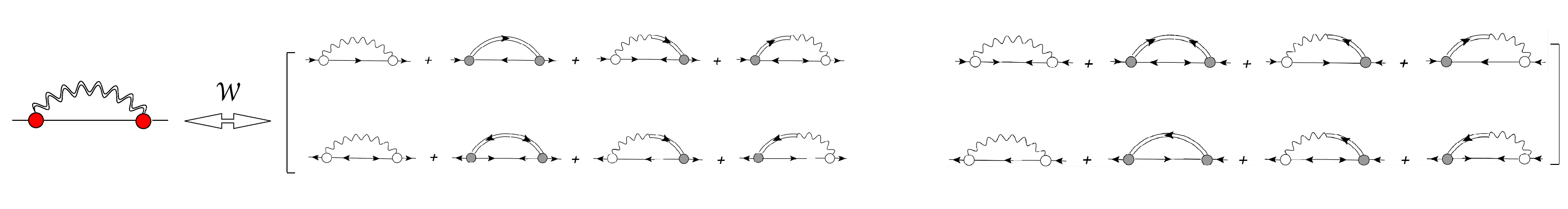}
\vspace{-1.0cm}
\end{center}
\caption{The superfluid qPVC self-energy in the single-particle (right) and quasiparticle (left) bases. The operation $\cal W$ stands for Bogolyubov's transformation.  Double wavy lines are introduced for the propagators of the superfluid phonons of the "unified" character in the quasiparticle basis, the associated filled (red) circles denote the respective combined phonon vertices and a single line without arrows is reserved for the quasiparticle propagator. The figure is adopted from Ref. \cite{Litvinova2022}.}
\label{SE_qvc}%
\end{figure*}

In analogy to the normal case, the components (\ref{Tsfdyn}) of the superfluid dynamical kernel can be treated in various approximations. In particular, one can apply the general cluster decomposition (\ref{CD}) to each of these components. However, in the superfluid ground state, more correlation functions will give non-vanishing contributions. As mentioned above, the approximation, where the many-body problem is truncated on the two-body level, i.e., the PVC approximation and its variants, is the most important one for applications to the regimes of intermediate coupling as it enables a good compromise between accuracy and feasibility. 
In the extension of the PVC approach to the superfluid, or quasiparticle, PVC dubbed as qPVC, the one-fermion propagators extend to the Gor'kov Green functions (\ref{GG}), and the two-fermion propagators include the contributions collected in Fig. \ref{QVC_map} in the diagrammatic representation.

After the corresponding algebra discussed in detail in Ref. \cite{Litvinova2021a}, in the qPVC approach the dynamical kernel $\Sigma^{r(+)}_{\nu\nu'}(\varepsilon)$ takes the form 
\bea
\Sigma^{r(+)}_{\nu\nu'}(\varepsilon) = \sum\limits_{\nu''\mu} \Bigl[ 
\frac{{\Gamma}^{(11)\mu}_{\nu\nu''}{\Gamma}^{(11)\mu\ast}_{\nu'\nu''}}{\varepsilon  - E_{\nu''} - \omega_{\mu} + i\delta} +
\frac{{\Gamma}^{(02)\mu\ast}_{\nu\nu''}{\Gamma}^{(02)\mu}_{\nu'\nu''}}{\varepsilon + E_{\nu''} + \omega_{\mu} - i\delta} \Bigl], \nonumber\\
\label{SEqp}
\eea
where the vertex functions $\Gamma^{(11)}$ and $\Gamma^{(02)}$ are defined as follows:
\bea
\Gamma^{(11)\mu}_{\nu\nu'} = \sum\limits_{12}\Bigl[ 
U^{\dagger}_{\nu 1}(g^{\mu}_{12}\eta^{\nu'}_2 + \gamma^{\mu(+)}_{12}\chi^{\nu'\ast}_2) - \nonumber \\
- V^{\dagger}_{\nu 1}((g^{\mu }_{12})^T\chi^{\nu'\ast}_2 + (\gamma^{\mu(-)}_{12})^T\eta^{\nu'}_2)\Bigr]
\label{Gamma11_qp}
\\
\Gamma^{(02)\mu}_{\nu\nu'} = -\sum\limits_{12}\Bigl[ 
V^{T}_{\nu 1}(g^{\mu}_{12}\eta^{\nu'}_2 + \gamma^{\mu(+)}_{12}\chi^{\nu'\ast}_2) - \nonumber \\
- U^{T}_{\nu 1}((g^{\mu }_{12})^T\chi^{\nu'\ast}_2 + (\gamma^{\mu(-)}_{12})^T\eta^{\nu'}_2)\Bigr].
\label{Gamma02_qp}
\eea
The $\eta = -$ component of $\Sigma^{r}_{\nu\nu'}(\varepsilon)$ has an analogous form, however, in practice the $\eta = -$ equation of the set (\ref{Dyson_qp}) is redundant as it has the same solution as its $\eta = +$ counterpart.  In the leading approximation, i.e., with the HFB values for the matrix elements $\eta^{\nu}_{i}$ and  $\chi^{\nu}_{i}$, Eqs. (\ref{Gamma11_qp},\ref{Gamma02_qp}) reduce to
\bea
\Gamma^{(11)\mu}_{\nu\nu'} = 
\Bigl[ 
U^{\dagger}g^{\mu}U + U^{\dagger}\gamma^{\mu(+)}V  \nonumber\\
- V^{\dagger}g^{\mu T}V - V^{\dagger}\gamma^{\mu(-)T}U\Bigr]_{\nu\nu'} \nonumber\\
\label{Gamma11_HFB}
\eea
\bea
\Gamma^{(02)\mu}_{\nu\nu'} =-\Bigl[ 
V^{T}g^{\mu}U+ V^{T}\gamma^{\mu(+)}V \nonumber\\
- U^{T}g^{\mu T}V - U^{T}\gamma^{\mu(-)T}U\Bigr]_{\nu\nu'}.\nonumber\\
\label{Gamma02_HFB}
\eea
In this way, one arrives at the compact form of the Gor'kov-Dyson equation (\ref{Dyson_qp}) in the qPVC approximation, where the dynamical kernel (\ref{SEqp}) has essentially the same form as in the non-superfluid case. The corresponding operation is shown in Fig. \ref{SE_qvc}.
All the complexity arising from the superfluidity is thus transferred to the qPVC vertices (\ref{Gamma11_qp},\ref{Gamma02_qp}). Essentially, in this framework both the normal and pairing phonons are unified in the superfluid phonons, whose vertices can be linked to the variations of the Hamiltonian of the Bogoliubov quasiparticles \cite{Litvinova2021a}.

\subsection{Response theory}
\label{response}

The EOMs for the two-fermion propagators defined by Eqs. (\ref{phresp},\ref{ppGF}) were obtained, for instance, in Refs. \cite{AdachiSchuck1989,DukelskyRoepkeSchuck1998} and, particularly, the PVC approximation was discussed in Refs. \cite{LitvinovaSchuck2019,Schuck2019,LitvinovaSchuck2020}.
The direct "superfluid" generalization of Eqs. (\ref{phresp},\ref{ppGF}) unifies these propagators, however, working in terms of Gor'kov Green functions (\ref{GG}) is quite cumbersome because of the quickly increasing number of components. It turns out that the formalism becomes more accessible if the EOM is derived in the quasiparticle space (\ref{Btrans}) from the beginning. While the detailed derivation was presented in Ref. \cite{Litvinova2022}, here we concentrate on the major building blocks of the general theory and concrete feasible approximations. 

In the quasiparticle basis, the Hamiltonian (\ref{Hamiltonian}) takes the following form \cite{RingSchuck1980}:
\bea
H = H^0 &+& \sum\limits_{\mu\nu}H^{11}_{\mu\nu}\alpha^{\dagger}_{\mu}\alpha_{\nu}  + \frac{1}{2}\sum\limits_{\mu\nu}\bigl(H^{20}_{\mu\nu}\alpha^{\dagger}_{\mu}\alpha^{\dagger}_{\nu} + \text{h.c.}\bigr) \nonumber\\
&+& \sum\limits_{\mu\mu'\nu\nu'}\bigl(H^{40}_{\mu\mu'\nu\nu'}\alpha^{\dagger}_{\mu}\alpha^{\dagger}_{\mu'}\alpha^{\dagger}_{\nu}\alpha^{\dagger}_{\nu'} + \text{h.c}\bigr) \nonumber\\
&+& \sum\limits_{\mu\mu'\nu\nu'}\bigl(H^{31}_{\mu\mu'\nu\nu'}\alpha^{\dagger}_{\mu}\alpha^{\dagger}_{\mu'}\alpha^{\dagger}_{\nu}\alpha_{\nu'} + \text{h.c}\bigr) 
\nonumber\\
&+& \frac{1}{4}\sum\limits_{\mu\mu'\nu\nu'}H^{22}_{\mu\mu'\nu\nu'}\alpha^{\dagger}_{\mu}\alpha^{\dagger}_{\mu'}\alpha_{\nu'}\alpha_{\nu}. 
\label{Hqua}
\eea
The upper indices in the matrix elements $H^{ij}_{\mu\nu}$ and  $H^{ij}_{\mu\nu\mu'\nu'}$ are associated with the numbers of creation and annihilation quasiparticle operators in the associated terms. The matrix elements $H^{ij}$ are listed, for instance, in \cite{RingSchuck1980}. The matrix $H^{20}$ vanishes at the stationary point defining the HFB equations, while the matrix elements of $H^{11}$ correspond to the quasiparticle energies. Thus, with $H^{11}_{\mu\nu} = \delta_{\mu\nu}E_{\mu}$, the Hamiltonian reads
 \cite{RingSchuck1980}:
 \be
 H = H^0 + \sum\limits_{\mu}E_{\mu}\alpha^{\dagger}_{\mu}\alpha_{\mu} + V,
 \label{Hqua1}
 \ee
where $V$ includes the remaining terms and has the meaning of the residual interaction:
 \bea
 V &=& 
 \sum\limits_{\mu\mu'\nu\nu'}\bigl(H^{40}_{\mu\mu'\nu\nu'}\alpha^{\dagger}_{\mu}\alpha^{\dagger}_{\mu'}\alpha^{\dagger}_{\nu}\alpha^{\dagger}_{\nu'} + \text{h.c}\bigr) \nonumber\\
&+& \sum\limits_{\mu\mu'\nu\nu'}\bigl(H^{31}_{\mu\mu'\nu\nu'}\alpha^{\dagger}_{\mu}\alpha^{\dagger}_{\mu'}\alpha^{\dagger}_{\nu}\alpha_{\nu'} + \text{h.c}\bigr) 
\nonumber\\
&+& \frac{1}{4}\sum\limits_{\mu\mu'\nu\nu'}H^{22}_{\mu\mu'\nu\nu'}\alpha^{\dagger}_{\mu}\alpha^{\dagger}_{\mu'}\alpha_{\nu'}\alpha_{\nu}. 
\label{Hqua2} 
\eea 

The definition of the superfluid response function to a sufficiently weak  external field $F$ can be deduced from the generic strength function:
\be
S(\omega) = \sum\limits_{n>0} \Bigl[ |\langle n|F^{\dagger}|0\rangle |^2\delta(\omega-\omega_n) - |\langle n|F|0\rangle |^2\delta(\omega+\omega_n)
\Bigr],
\label{SF}
\ee
where the summation runs over all the formally exact excited states $|n\rangle$. 
The generic one-body operator $F$ in terms of the quasiparticle fields reads:
\bea
F = \frac{1}{2}\sum\limits_{\mu\mu'} \Bigl(F^{20}_{\mu\mu'}\alpha^{\dagger}_{\mu}\alpha^{\dagger}_{\mu'} + 
F^{02}_{\mu\mu'}\alpha_{\mu'}\alpha_{\mu} \Bigr)\nonumber\\
F^{\dagger} = \frac{1}{2}\sum\limits_{\mu\mu'} \Bigl(F^{20\ast}_{\mu\mu'}\alpha_{\mu'}\alpha_{\mu} +
F^{02\ast}_{\mu\mu'}\alpha^{\dagger}_{\mu}\alpha^{\dagger}_{\mu'}  
\Bigr),
\label{Fext}
\eea
following Bogolyubov's transformation of the second-quantized form of $F$. The full composition in the quasiparticle basis contains formally also $F^{11}$ terms, however, their contribution vanishes in the leading approximations to the superfluid response \cite{Avogadro2011}. Subleading contributions will be considered elsewhere. Eq. (\ref{SF}) can be transformed as follows:
\bea
S(\omega) &=& -\frac{1}{\pi}\lim\limits_{\Delta \to 0} \text{Im} \Pi(\omega),\nonumber\\
\label{SFDelta} 
\Pi(\omega) &=&  \frac{1}{4}\sum\limits_{\mu\mu'\nu\nu'}
\left(\begin{array}{cc} F^{02}_{\mu\mu'} & F^{20}_{\mu\mu'} \end{array}\right){\hat{\cal R}}_{\mu\mu'\nu\nu'}(\omega+i\Delta)\left(\begin{array}{c} F^{02\ast}_{\nu\nu'} \\  \\ F^{20\ast}_{\nu\nu'} \end{array}\right),\nonumber\\
\label{Polar}
\eea
where the matrix of the response function reads:
\bea
{\hat{\cal R}}_{\mu\mu'\nu\nu'}(\omega) = \nonumber \\
= \sum\limits_{n>0} \left(\begin{array}{c} {\cal X}^{n}_{\mu\mu'} \\ {\cal Y}^{n}_{\mu\mu'} \end{array}\right)
\frac{1}{\omega - \omega_n + i\delta}\left(\begin{array}{cc} {\cal X}^{n\ast}_{\nu\nu'} & {\cal Y}^{n\ast}_{\nu\nu'} \end{array}\right)\nonumber \\
- \sum\limits_{n>0} \left(\begin{array}{c} {\cal Y}^{n\ast}_{\mu\mu'} \\ {\cal X}^{n\ast}_{\mu\mu'} \end{array}\right)
\frac{1}{\omega + \omega_n - i\delta}\left(\begin{array}{cc} {\cal Y}^{n}_{\nu\nu'} & {\cal X}^{n}_{\nu\nu'} \end{array}\right),\nonumber \\
\label{Romega}
\eea
with the matrix elements
\be
{\cal X}^{n}_{\mu\mu'} = \langle 0|\alpha_{\mu'}\alpha_{\mu}|n\rangle \ \ \ \ \ \ \ \ \ \ {\cal Y}^{n}_{\mu\mu'} = \langle 0|\alpha^{\dagger}_{\mu}\alpha^{\dagger}_{\mu'}|n\rangle .
\label{XY}
\ee
In terms of the time-dependent field operators, it can be thus defined as 
\bea
{\hat{\cal R}}_{\mu\mu'\nu\nu'} (t-t') = \nonumber
\\
= -i\langle T\left(\begin{array}{cc}
(\alpha_{\mu'}\alpha_{\mu})(t)(\alpha^{\dagger}_{\nu}\alpha^{\dagger}_{\nu'})(t') & 
(\alpha_{\mu'}\alpha_{\mu})(t)(\alpha_{\nu'}\alpha_{\nu})(t') \\
(\alpha^{\dagger}_{\mu}\alpha^{\dagger}_{\mu'})(t)(\alpha^{\dagger}_{\nu}\alpha^{\dagger}_{\nu'})(t') &
(\alpha^{\dagger}_{\mu}\alpha^{\dagger}_{\mu'})(t)(\alpha_{\nu'}\alpha_{\nu})(t')
\end{array}\right)\rangle 
\nonumber\\
= -i
\langle T\left(\begin{array}{cc}
A_{\mu\mu'}(t)A^{\dagger}_{\nu\nu'}(t')  &
A_{\mu\mu'}(t)A_{\nu\nu'}(t') \\
A^{\dagger}_{\mu\mu'}(t)A^{\dagger}_{\nu\nu'}(t') &
A^{\dagger}_{\mu\mu'}(t)A_{\nu\nu'}(t')
\end{array}\right)\rangle ,\nonumber\\
\label{RtA}
\eea
where we introduce the time-dependent operator products in the Heisenberg picture:
\bea
A_{\mu\mu'}(t) = (\alpha_{\mu'}\alpha_{\mu})(t) = e^{iHt}\alpha_{\mu'}\alpha_{\mu}e^{-iHt} \nonumber \\ 
A^{\dagger}_{\nu\nu'}(t) = (\alpha^{\dagger}_{\nu}\alpha^{\dagger}_{\nu'})(t) = 
e^{iHt}\alpha^{\dagger}_{\nu}\alpha^{\dagger}_{\nu'}e^{-iHt}.
\eea 

The EOM for the superfluid response (\ref{RtA}) is generated by the same technique. Differentiating Eq. (\ref{RtA}) sequentially with respect to $t$ and $t'$ and performing the Fourier transformation, one obtains
\bea
{\hat{\cal R}}_{\mu\mu'\nu\nu'}(\omega) = {\hat{\cal R}}^0_{\mu\mu'\nu\nu'}(\omega) \nonumber\\
+ \frac{1}{4}{\hat{\cal R}}^0_{\mu\mu'\gamma\gamma'}(\omega){\hat{\cal K}}_{\gamma\gamma'\delta\delta'}(\omega){\hat{\cal R}}_{\delta\delta'\nu\nu'}(\omega),
\label{BSDE}
\eea
which has the form of the Bethe-Salpeter-Dyson equation, but with the $2\times 2$ matrix structure in the quasiparticle basis. The free response is meanwhile defined as
\be
{\hat{\cal R}}^0_{\mu\mu'\nu\nu'}(\omega) = \left[\omega\right. - \left.{\hat\sigma}_3E_{\mu\mu'}\right]^{-1}{\hat{\cal N}}_{\mu\mu'\nu\nu'},
\ee
with
\be
E_{\mu\mu'} = E_{\mu} + E_{\mu'}, \ \ \ \ \ \ \ \ \ \ \ \ \ \ \ {\hat\sigma}_3 = \left(\begin{array}{cc} 1& 0 \\ 0 & -1 \end{array}\right)
\ee 
and the norm matrix ${\hat{\cal N}}_{\mu\mu'\nu\nu'}$ specified below.
The interaction kernel is given by
\bea
{\hat{\cal K}}^0_{\gamma\gamma'\delta\delta'} = \frac{1}{4}{\hat{\cal N}}^{-1}_{\gamma\gamma'\eta\eta'}
{\hat{\cal T}}^{0}_{\eta\eta'\rho\rho'} 
{\hat{\cal N}}^{-1}_{\rho\rho'\delta\delta'} \nonumber\\
{\hat{\cal K}}^r_{\gamma\gamma'\delta\delta'}(\omega) = \frac{1}{4}\left[{\hat{\cal N}}^{-1}_{\gamma\gamma'\eta\eta'}
{\hat{\cal T}}^{r}_{\eta\eta'\rho\rho'}(\omega) 
{\hat{\cal N}}^{-1}_{\rho\rho'\delta\delta'}\right]^{irr}.
\eea
with the static and dynamical ${\hat{\cal T}}$-matrices in the quasiparticle space:
\be
{\hat{\cal T}}^{0}_{\mu\mu'\nu\nu'} = -\langle\left(\begin{array}{cc}\left[[V,A_{\mu\mu'}],A^{\dagger}_{\nu\nu'}\right]  & \Bigl[[V,A_{\mu\mu'}],A_{\nu\nu'}\Bigr] 
\\
\\
\left[[V,A^{\dagger}_{\mu\mu'}],A^{\dagger}_{\nu\nu'}\right]  &
\left[[V,A^{\dagger}_{\mu\mu'}],A_{\nu\nu'}\right]
\end{array}\right)\rangle
\label{T02}
\ee
\bea
{\hat{\cal T}}^{r}_{\mu\mu'\nu\nu'}(t-t') = i\times\ \ \ \ \ \ \ \ \ \ \ \ \ \ \ \ \ \ \ \ \ \ \ \ \ \ \ \ \ \ \ \ \ \ \ \ \ \ \ \ \ \ \ \ \ \
\nonumber\\ \times
\langle T\left(\begin{array}{cc}[V,A_{\mu\mu'}](t)[V,A^{\dagger}_{\nu\nu'}](t')  & [V,A_{\mu\mu'}](t)[V,A_{\nu\nu'}](t')
\\
\left[V,A^{\dagger}_{\mu\mu'}\right](t)[V,A^{\dagger}_{\nu\nu'}](t')  &
[V,A^{\dagger}_{\mu\mu'}](t)[V,A_{\nu\nu'}](t')
\end{array}\right)\rangle . \nonumber\\
\label{Tr2}
\eea
The norm matrix ${\hat{\cal N}}_{\mu\mu'\nu\nu'}$ also acquires an extended form and reads:
\be
{\hat{\cal N}}_{\mu\mu'\nu\nu'} = \langle\left(\begin{array}{cc}[A_{\mu\mu'},A^{\dagger}_{\nu\nu'}]  & 0
\\
0  &
[A^{\dagger}_{\mu\mu'},A_{\nu\nu'}]
\end{array}\right)\rangle ,
\label{norm}
\ee
with the inverse introduced according to the identity:
\bea
\frac{1}{2}\sum_{\delta\delta'}{\hat{\cal N}}^{-1}_{\mu\mu'\delta\delta'}{\hat{\cal N}}_{\delta\delta'\nu\nu'} = \delta_{\mu\mu'\nu\nu'} =
 \delta_{\mu\nu}\delta_{\mu'\nu'} - \delta_{\mu\nu'}\delta_{\mu'\nu}.\nonumber\\
\eea

So far the theory is still very general and, in order to proceed, evaluation of the double commutators of Eq. (\ref{T02}) and the commutator products of Eq. (\ref{Tr2}) is required. 
The ab-initio form of the static kernel (\ref{T02}) is the unification of the $ph$ and $pp$ static kernels discussed in Refs. \cite{SchuckTohyama2016a,LitvinovaSchuck2019,Schuck2019,LitvinovaSchuck2020,Schuck2021}. In addition to the pure contributions from the bare fermionic interaction, these kernels contain the terms with contractions of the interaction with the correlated parts of the two-body fermionic densities. The latter should be generalized to the superfluid two-body densities and include feedback from the dynamical kernel, which will be considered elsewhere. 
If the ground state is confined by the HFB approximation, the superfluid static kernel simplifies to the well-known kernel of the quasiparticle random phase approximation (QRPA), with the expectation values of the commutators 
\bea
\langle\text{HFB}|\left[[V,A_{\mu\mu'}],A^{\dagger}_{\nu\nu'}\right]|\text{HFB}\rangle 
&=& -H^{22}_{\mu\mu'\nu\nu'},\nonumber
\\
\langle\text{HFB}|\left[[V,A_{\mu\mu'}],A_{\nu\nu'}\right]|\text{HFB}\rangle &=& 4!H^{40}_{\mu\mu'\nu\nu'},\nonumber
\\
{\hat{\cal N}}_{\mu\mu'\nu\nu'} &=& {\hat{\sigma}}_3\delta_{\mu\mu'\nu\nu'},
\nonumber\\
\label{ABN}
\eea
while the remaining matrix elements of Eq. (\ref{T02}) can be obtained by Hermitian conjugation. 
\begin{figure*}
\begin{center}
\includegraphics[scale=0.70]{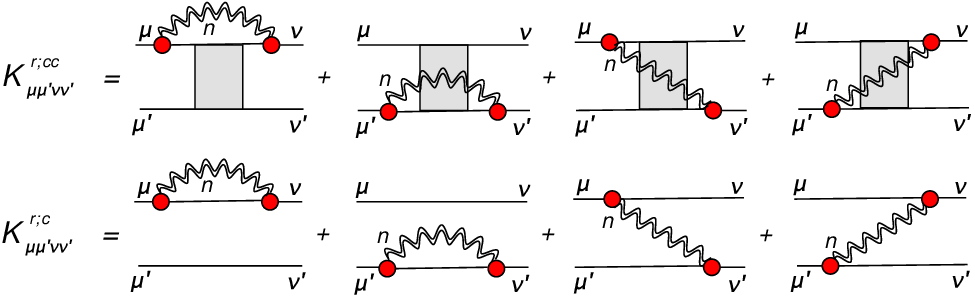}
\end{center}
\caption{The leading approximations to the superfluid dynamical kernel taking into account the qPVC effects. The shaded rectangular block denotes the generic two-quasiparticle correlation function (\ref{RtA}). Top: the kernel with two two-quasiparticle correlation functions (\ref{Kr11cc}); bottom: the kernel with one two-quasiparticle correlation function (\ref{Kr11c}).}
\label{Kr_qvc}%
\end{figure*}

The components of the dynamical kernel defined by the time-dependent commutator products of Eq. (\ref{Tr2}) can be evaluated similarly. The diagonal components provide the dominant contribution, while the non-diagonal ones contribute via complex ground-state correlations. The latter will be neglected in this work.
Furthermore, the diagonal components are connected by the relationship
\be
{\cal K}^{r[22]}_{\mu\mu'\nu\nu'}(\tau) = {\cal K}^{r[11]}_{\nu\nu'\mu\mu'}(-\tau),
\ee
so that it is sufficient to explicitly obtain ${\cal K}^{r[11]}_{\nu\nu'\mu\mu'}$. Its $T$-matrix counterpart
\be
{\cal T}^{r[11]}_{\mu\mu'\nu\nu'}(t-t') = i\langle T[V,A_{\mu\mu'}](t)[V,A^{\dagger}_{\nu\nu'}](t') 
\rangle
\label{Tr11}
\ee
generates a product of eight quasiparticle operators, four at time $t$ and four at time $t'$, i.e., fully correlated two-times four-quasiparticle propagator, contracted with two matrix elements of the residual interaction. The appearance of such a propagator in the dynamical kernel signals about generating of a hierarchy of coupled EOMs for growing-rank propagators also in the two-fermionic sector.  Again, various approximations of increasing accuracy constructed by a factorization procedure are possible, and here we narrow the discussion to the factorizations, which keep all the possible contributions with two-fermion propagators (\ref{RtA}). After dropping the complex ground state correlation contributions, the ${\cal K}^{r[11]}$ component takes the form
\bea
{\cal K}^{r[11]cc}_{\mu\mu'\nu\nu'}(\omega) &=& 
\sum\limits_{\gamma\delta nm}\Bigl[\frac{\Gamma^{(11)n}_{\mu\gamma}{\cal X}^{m}_{\mu'\gamma}{\cal X}^{m\ast}_{\nu'\delta}\Gamma^{(11)n\ast}_{\nu\delta}}{\omega - \omega_{nm} + i\delta} \nonumber\\ &-& 
\frac{\Gamma^{(11)n\ast}_{\gamma\mu}{\cal Y}^{m\ast}_{\mu'\gamma}{\cal Y}^{m}_{\nu'\delta}\Gamma^{(11)n}_{\delta\nu}}{\omega + \omega_{nm} - i\delta}\Bigr]
- \cal{AS},
\label{Kr11cc}
\eea
where $\cal{AS}$ includes all the antisymmetrizations and the additional upper index "cc" indicates the approximation, where two fully correlated two-fermion propagators are retained. The diagrammatic interpretation of Eq. (\ref{Kr11cc}) is illustrated in Fig. \ref{Kr_qvc}.
One can notice that, although two two-fermion CFs are figuring in this kernel, only one of them forms a phonon, because only two matrix elements of the NN interaction are entering the initial expression (\ref{Tr2}), and thus only one CF contracted with these matrix elements can be mapped to the PVC according to Eqs. (\ref{mappingph},\ref{mappingpp}). This is at variance with the QPM, see, for instance, the QPM kernels analyzed in Ref. \cite{Malov1985}, which contain the two-phonon and, in principle, further multiphonon contributions. These contributions are practically postulated in the ansatz of the excited state wave function, whose coefficients are then found variationally. It is not straightforwardly clear, therefore, if the $n$-phonon QPM with $n>1$ can be obtained as a cluster approximation to the dynamical kernel (\ref{Tr2}). This may be possible in further approximation to the phonons where they are confined by the (Q)RPA as, in this case, the quasiparticle pair operators can be expressed via the phonon operators; however, we leave this as a conjecture here for a future more accurate investigation. 

We note also that the two-quasiparticle CFs and phonons appearing in the dynamical kernel (\ref{Kr11cc}) are formally exact and, therefore, are not associated with any partial resummations or perturbative expansions. This indicates that the cluster approaches discussed in this section can include arbitrarily complex 2n-quasiparticle configurations. However, approximations can always be applied to calculations of the phonons.

Finally, by relaxing the correlations in the intermediate two-quasiparticle propagator, which is not associated with a phonon, in Eq. (\ref{Kr11cc}), one obtains its qPVC-NFT approximation:
\bea
{\cal K}^{r[11]c}_{\mu\mu'\nu\nu'}(\omega) = 
\Bigl\{\Bigl[ \delta_{\mu'\nu'}
\sum\limits_{\gamma n}\frac{\Gamma^{(11)n}_{\mu\gamma}\Gamma^{(11)n\ast}_{\nu\gamma}}{\omega - \omega_{n} - E_{\mu'} - E_{\gamma}}  \nonumber\\ - 
\sum\limits_{n}\frac{\Gamma^{(11)n}_{\mu\nu'}\Gamma^{(11)n\ast}_{\nu\mu'}}{\omega - \omega_{n} - E_{\mu'} - E_{\nu'}}\Bigr]
- \Bigl[ \mu\leftrightarrow\mu'\Bigr]\Bigr\} - \Bigl\{ \nu\leftrightarrow\nu'\Bigr\},\nonumber\\
\label{Kr11AAa_qPVC_0}
\eea
where we indicated by the index "c" that only one two-quasiparticle correlation function is retained in the dynamical kernel.
To bring this kernel to the form, which corresponds to the superfluid generalization of the NFT, one can recast Eq. (\ref{Kr11AAa_qPVC_0}) by performing the explicit antisymmetrizations and rearranging the resulting terms as follows:
\bea
{\cal K}^{r[11]c}_{\mu\mu'\nu\nu'}(\omega) =  \ \ \ \ \ \ \ \ \ \ \ \ \ \ \ \ \ \ \ \  \nonumber\\
= \Bigl[ \delta_{\mu'\nu'}
\sum\limits_{\gamma n}\frac{\Gamma^{(11)n}_{\mu\gamma}\Gamma^{(11)n\ast}_{\nu\gamma}}{\omega - \omega_{n} - E_{\mu'\gamma}}  
+\delta_{\mu\nu}
\sum\limits_{\gamma n}\frac{\Gamma^{(11)n}_{\mu'\gamma}\Gamma^{(11)n\ast}_{\nu'\gamma}}{\omega - \omega_{n} - E_{\mu\gamma}}
\nonumber\\ + 
\sum\limits_{n}\frac{\Gamma^{(11)n}_{\mu\nu}\Gamma^{(11)n\ast}_{\nu'\mu'}}{\omega - \omega_{n} - E_{\mu'\nu}}
+ \sum\limits_{n}\frac{\Gamma^{(11)n}_{\mu'\nu'}\Gamma^{(11)n\ast}_{\nu\mu}}{\omega - \omega_{n} - E_{\mu\nu'}}\Bigr] \nonumber\\
- \Bigl[ \nu\leftrightarrow\nu'\Bigr].\ \ \ \ \ \ \ \ \ \ \ \ 
\label{Kr11c}
\eea

This form of the dynamical kernel can be further simplified if the pairing correlations are approximated by the BCS theory, see for instance, Ref. \cite{Zelevinsky2017}.
It is essentially an analog of the resonant kernel obtained in phenomenological approaches of the NFT \cite{Niu2016} and quasiparticle time blocking approximation \cite{Tselyaev2007,LitvinovaTselyaev2007}. Finally, we note that on the way to Eqs. (\ref{Kr11cc}-\ref{Kr11c}) a number of correlations were neglected in the versions ${\cal K}^{r[11]cc}$ and ${\cal K}^{r[11]c}$ of the kernel, which are considered to be the leading, sometimes called resonant, approximations.  The major subleading contributions are then associated with the $\langle 0|\alpha^{\dagger}_{\mu}\alpha_{\mu'}|n\rangle$ amplitudes, the terms of the residual interaction other than $H^{31}$, and the off-diagonal ${\cal K}^{r[12]}$ and ${\cal K}^{r[21]}$ contributions. They can be straightforwardly included due to the universality and completeness of the presented framework, which thus offers a number of extensions beyond the qPVC approaches ${\cal K}^{r[11]cc}$ and ${\cal K}^{r[11]c}$ given explicitly in this section. 

\subsection{Self-consistent implementations of the qPVC approaches for nuclear response}

The simplest resonant qPVC dynamical kernel ${\cal K}^{r[11]c}$ has been a subject of self-consistent implementations based on the effective in-medium NN interactions derived from the DFT. The applications to the nuclear response include, for instance, the ones with the zero-range phenomenological Skyrme interaction \cite{Tselyaev2018,Lyutorovich2018,Niu2016,Niu2018}. The self-consistency in this context means that (i) the static kernels of the one-fermion and two-fermion EOMs are approximated by the first and second variational derivatives of the EDF, respectively, (ii) the phonons' characteristics are obtained with the same interaction, and (iii) the subtraction of the static limit of the dynamical kernel \cite{Tselyaev2013} is applied to eliminate the double counting of its admixture to the effective interaction.    

The fully self-consistent qPVC approaches to the nuclear response with the more fundamental background of the relativistic meson-nucleon Lagrangian are available since Ref. \cite{LitvinovaRingTselyaev2008} under the relativistic time blocking approximation (RQTBA) in the more general context of the relativistic nuclear field theory (RNFT) and include calculations for both the neutral \cite{LitvinovaLoensLangankeEtAl2009,LitvinovaRingTselyaev2010,EndresLitvinovaSavranEtAl2010,MassarczykSchwengnerDoenauEtAl2012,LanzaVitturiLitvinovaEtAl2014,PoltoratskaFearickKrumbholzEtAl2014,NegiWiedekingLanzaEtAl2016,EgorovaLitvinova2016,Carter2022,Litvinova2023} and charge-exchange \cite{RobinLitvinova2016,Scott2017,RobinLitvinova2018,Robin2019} excitations. Important improvements due to the inclusion of the qPVC dynamical kernel have been found 
already in the leading approximation (\ref{Kr11c}). The widths of the giant resonances, the isospin character of the low-energy soft modes and overall strength distributions, beta decay, nuclear compressibility, and other nuclear structure properties were described in a single framework with universal parameters across the nuclear chart. The complete self-consistency, the covariant nature of the RNFT, and its background in the microscopic NN interaction, only slightly adjusted to the nuclear medium, provide a good balance of fundamentality and accuracy, making this theory most predictive, transferrable across the energy scales and systematically improvable. The latter two features were enabled after the recent completion of the ab-initio EOM qPVC framework \cite{LitvinovaSchuck2019,LitvinovaSchuck2020,Litvinova2021a,Litvinova2022} outlined in the previous sections. Other recent developments extend the RNFT to finite temperature \cite{LitvinovaWibowo2018,LitvinovaWibowo2019,LitvinovaRobinWibowo2020,Litvinova2021b}, making it the theory of choice for predictive astrophysical applications, and include correlations of higher complexity, \cite{Litvinova2015,Robin2019,LitvinovaSchuck2019}, heading toward spectroscopically accurate calculations in large model spaces.

As the time blocking \cite{Tselyaev2007} applied to the four-point correlation functions is not needed for the two-point response, the qPVC approaches presented in this work are dubbed as relativistic EOM (REOM$^n$) with the upper index indicating the maximal configuration complexity of the dynamical kernel. 

\section{Nuclear response in RNFT framework}

The response function is commonly actualized via its convolution with the operators, which are associated with external probes of the system under study.
The resulting quantity which describes the observed spectra is the strength function, which for the given external field operator $F$ can be expressed as
\be
S_F(\omega) = -\frac{1}{\pi}\lim_{\Delta\to 0}\Im\sum\limits_{121'2'}F_{12}R_{12,1'2'}(\omega+i\Delta)F^{\ast}_{1'2'},
\label{SFF}
\ee
while its form in the quasiparticle basis is given in subsection \ref{response}.
In this work, we consider the response to the electromagnetic dipole and quadrupole operators, respectively,
\bea
F^{(\text{EME1})}_{1M} &=& \frac{eN}{A}\sum\limits_{i=1}^Z r_iY_{1M}({\hat{\bf r}}_i) - \frac{eZ}{A}\sum\limits_{i=1}^N r_iY_{1M}({\hat{\bf r}}_i),  \nonumber \\
F^{(\text{EME2})}_{2M} &=& e\sum\limits_{i=1}^Z r^2_iY_{2M}({\hat{\bf r}}_i) 
\label{opE1E2}
\eea 
where $Z$ and $N$ are the numbers of protons and neutrons, respectively, $A = N+Z$, and $e$ is the proton charge, which is taken $e = 1$. The sums in Eq. (\ref{opE1E2}) are performed over the corresponding nucleonic degrees of freedom. 

\begin{figure}
\begin{center}
\includegraphics[scale=0.38]{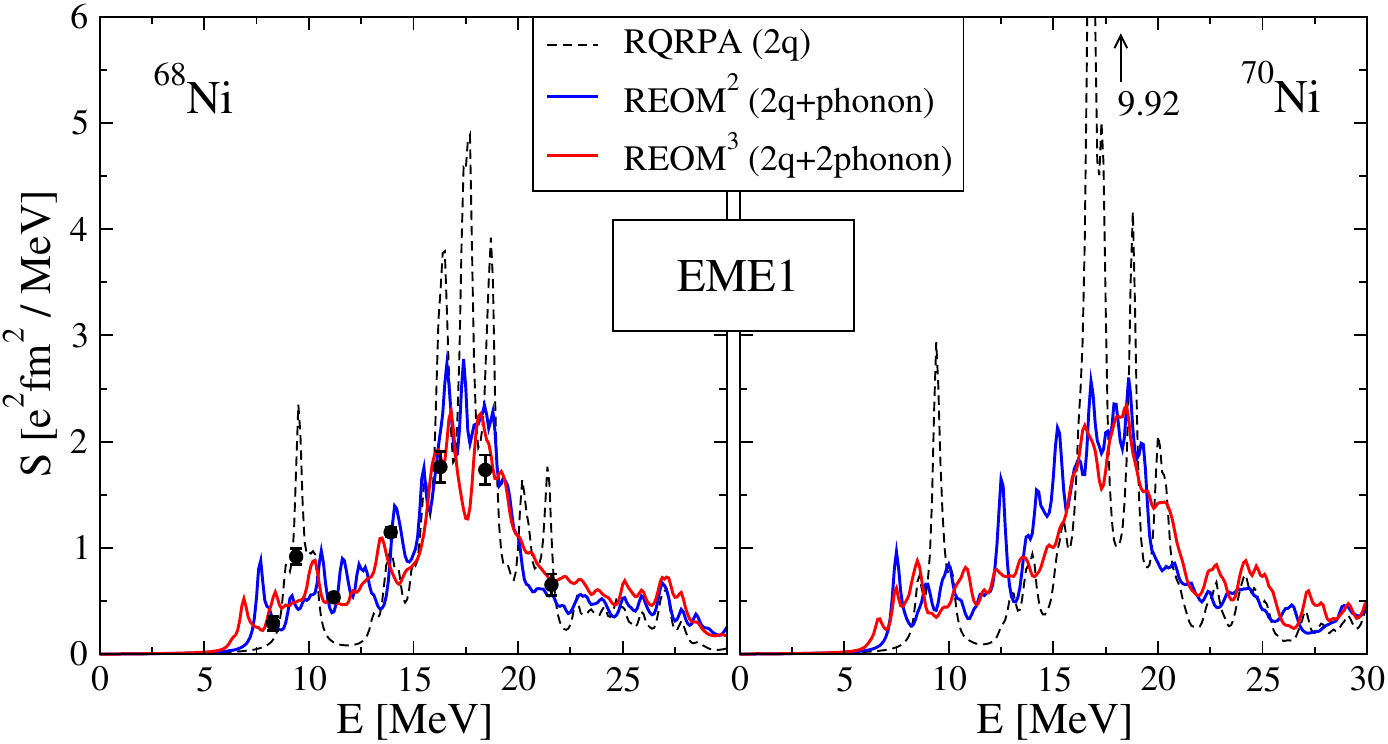}
\end{center}
\caption{Electromagnetic dipole response of $^{68,70}$Ni in approximations of growing complexity derived within the EOM framework. The experimental data for $^{68}$Ni are adopted from Ref. \cite{Rossi2013}, see text for details.}
\label{Ni-GDR}%
\end{figure}
\begin{figure}
\begin{center}
\includegraphics[scale=0.38]{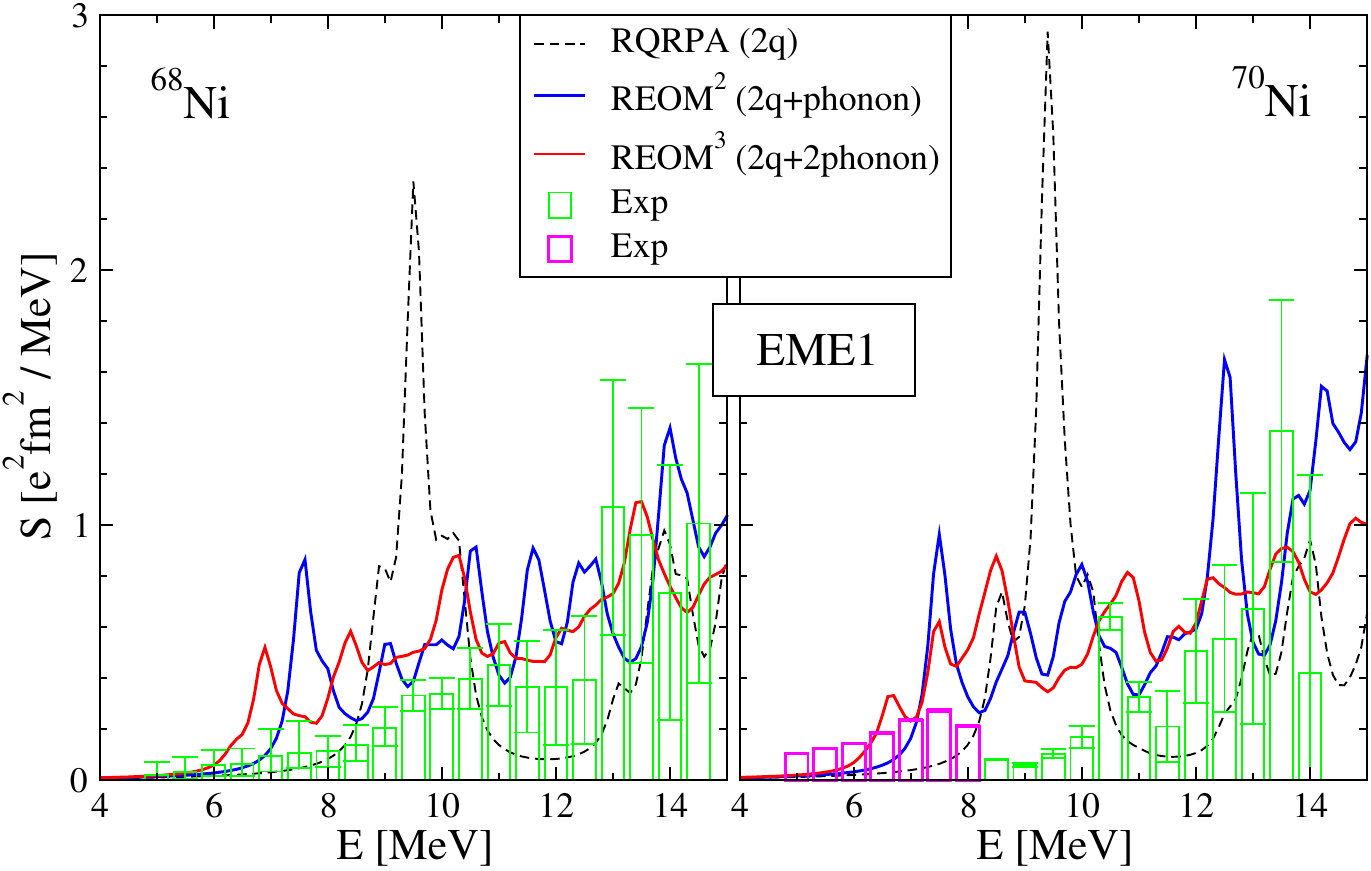}
\end{center}
\caption{Low-energy fraction of the electromagnetic dipole response of $^{68,70}$Ni obtained in the same approximations as in Fig. \ref{Ni-GDR} compared to data of Ref. \cite{Wieland2018}.}
\label{Ni-PDR}%
\end{figure}

Figs. \ref{Ni-GDR} and \ref{Ni-PDR} illustrate the RNFT calculations of the dipole response of the two neutron-rich nickel isotopes $^{68,70}$Ni. The theoretical results are obtained in the three relativistic approaches: (i) Relativistic QRPA (RQRPA) \cite{PaarRingNiksicEtAl2003} confined by the $2q$ configurations, (ii) relativistic EOM confined by correlated $2p2h$, or four-quasiparticle, configurations $2q\otimes phonon$, employing the dynamical kernel (\ref{Kr11c}) illustrated in the lower part of Fig. \ref{Kr_qvc} and dubbed as REOM$^2$, and (iii)  relativistic EOM confined by correlated $3p3h$, or six-quasiparticle, configurations $2q\otimes 2phonon$, employing the dynamical kernel (\ref{Kr11cc}) illustrated in the upper part of Fig. \ref{Kr_qvc} and dubbed as REOM$^3$. 
In the latter approximation, the $2q\otimes 2phonon$ configuration complexity is achieved by recycling the $2q\otimes phonon$ correlated propagators in the kernel 
of Eq. (\ref{Kr11cc}).
The following multi-step calculation scheme is implemented: 
\begin{itemize}
\item
The closed set of the relativistic mean field (RMF) Hartree-Bogoliubov equations \cite{SerotWalecka1986a,Ring1996,VretenarAfanasjevLalazissisEtAl2005} is solved with the NL3 effective interaction following Refs. \cite{BogutaBodmer1977,Lalazissis1997}, with the monopole pairing forces adjusted to reproduce empirical pairing gaps. The resulting single-particle Dirac spinors and single-nucleon energies form the working basis for subsequent calculations. No further parameters are introduced.
\item
The RQRPA equation \cite{PaarRingNiksicEtAl2003}, which is equivalent to Eq. (\ref{BSDE}) with the only static kernel (\ref{T02}) approximated by Eqs. (\ref{ABN}) with the effective interaction matrix elements, is solved to obtain the phonon vertices $\Gamma^n$ and their frequencies $\omega_{n}$. 
The phonons with the $J_{n}^{\pi_{n}}$ = 2$^+$, 3$^-$, 4$^+$, 5$^-$, 6$^+$ and frequencies $\omega_{n} \leq$ 15 MeV coupled
to the RMF quasiparticle states form the $2q\otimes phonon$ configuration space for the qPVC kernels. The phonon space was additionally truncated: the modes with the values of the reduced transition probabilities $B(EL)$ less than 5\% of the maximal one (for each $J_{n}^{\pi_{n}}$) were neglected. These are the common truncation criteria for the qPVC models based on the RMF, see, for instance, \cite{Litvinova2016}, where a convergence study was presented. The convergence is further reinforced by the subtraction procedure \cite{Tselyaev2013}.
\item 
{\it Additional step for the REOM$^{\ 3}$ approach}: Eq. (\ref{BSDE}) for the response function is solved in the truncated configuration space, which includes excitations within the energy window of interest 0-25 MeV, for spins and parities $J^{\pi}$ = 0$^{\pm}$ - 6$^{\pm}$. As in Ref. \cite{LitvinovaSchuck2019}, the static part of the kernel is neglected in the internal correlation functions as it does not induce fragmentation. The resulting response functions are inserted into the kernel (\ref{Kr11cc}).
\item 
The obtained dynamical kernels  (\ref{Kr11c}) REOM$^{2}$ and (\ref{Kr11cc}) REOM$^{3}$ are used in solving Eq. (\ref{BSDE}) for the main channels under study $J^{\pi} = 1^-$ and $J^{\pi} = 2^+$. The subtraction \cite{Tselyaev2013} is applied in both REOM$^{2}$ and REOM$^{3}$ calculations to eliminate the double counting of the qPVC from the effective interaction.
\item
Finally, the strength functions are found according to Eq. (\ref{Polar}).
 \end{itemize}

Further details of the calculation scheme can be found in Ref. \cite{LitvinovaSchuck2019}, where the REOM$^3$ was presented for the first time and applied to the dipole response of calcium isotopes. Here we focus on the dipole transitions of the neutron-rich unstable nickel isotopes, which are extensively studied both theoretically and experimentally and which are of special significance for astrophysics. 

In the left panel of Fig. \ref{Ni-GDR} the dipole strength distributions in $^{68}$Ni obtained in the three approaches with the same parameter set and with $\Delta = 200$ keV illustrate the hierarchy of approximations of growing complexity.
Experimental data from Ref. \cite{Rossi2013} are plotted for comparison after scaling to accommodate the enhancement of the Thomas-Reiche-Kuhn sum rule. Remarkably, the REOM$^2$, which includes the leading effects of emergent collectivity, provides a major improvement of the description. The REOM$^3$ including more complex qPVC correlations in the dynamical kernel further refines the strength distribution, but the overall effect on the smeared strength is less drastic. Similar results are obtained for $^{70}$Ni, for which the experimental data are available only at low energies, see Fig. \ref{Ni-PDR} and discussion below. 
%
\begin{figure*}
\begin{center}
\vspace{-0.3cm}
\includegraphics[scale=0.52]{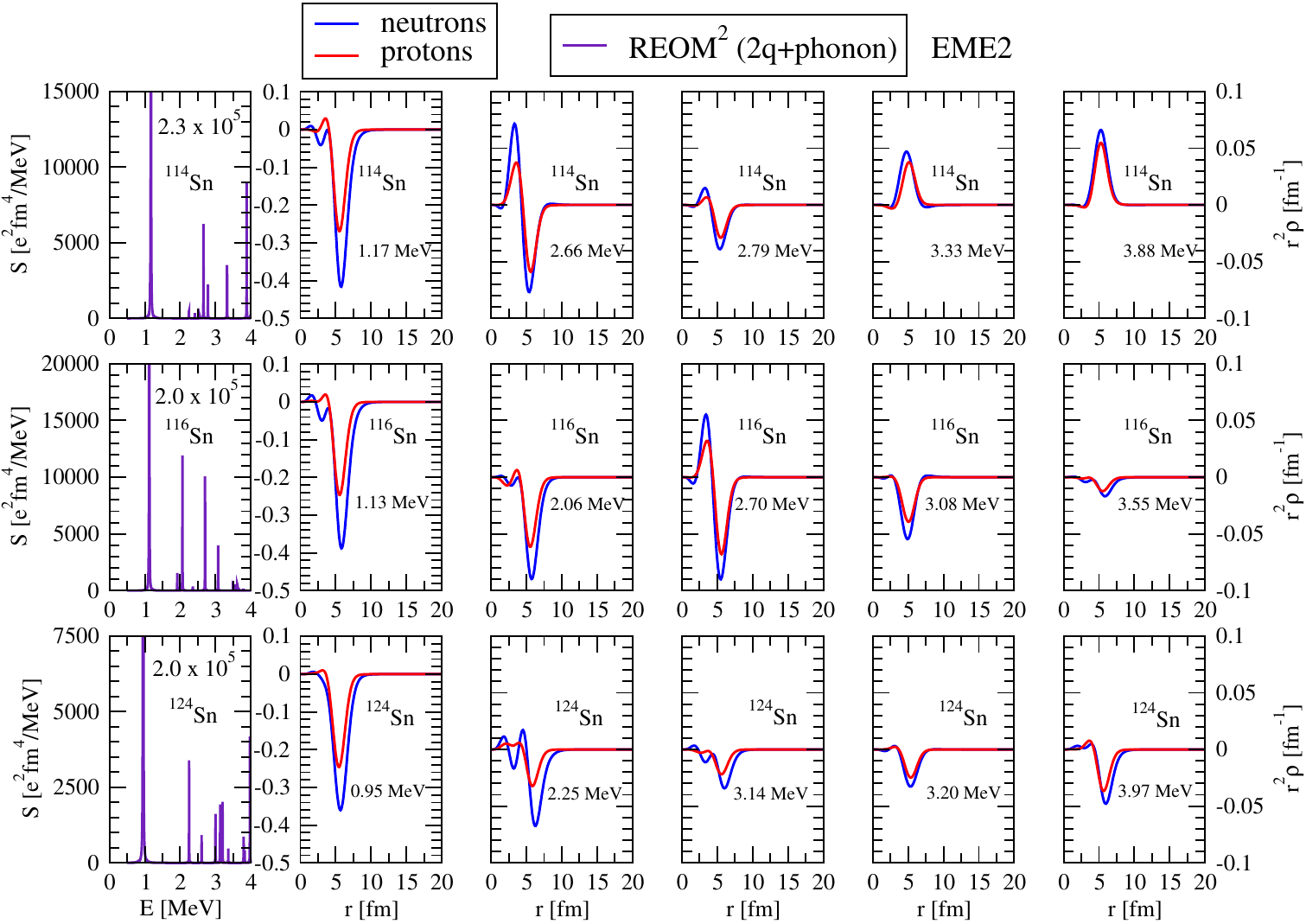}
\end{center}
\vspace{-0.6cm}
\caption{Quadrupole strength below 4 MeV in $^{114,116,124}$Sn isotopes calculated in REOM$^2$ approach (left panels) and the proton and neutron transition densities for the most prominent peaks.}
\label{quad_1}%
\end{figure*}

Calculations in the REOM$^3$ approach are significantly heavier than those in REOM$^2$ because a large, ideally complete, set of the internal CFs enters the doubly correlated kernel of REOM$^3$. The set of phonons coupled to those CFs in the kernel of Eq. (\ref{Kr11cc}) is also formally complete, however, the CFs in the phonons enter in the form of contraction with the interaction matrix elements, i.e., with the phonons' vertices $\Gamma^n$. As it follows from numerous previous RQTBA studies, these vertices behave as emergent order parameters, which enables reasonably accurate truncation schemes of the phonon model subspaces. Moreover, as the most significant phonons are well reproduced within RQRPA, reiterating their CFs does not bring significant improvements. However, reiterating and keeping a sufficiently complete set of the internal non-contracted CFs appears to be quantitatively important.
This is the major factor that makes large-scale studies difficult, however, simple parallelization algorithms can be implemented to improve the time scaling of the REOM$^3$ approach. 

The cases of $^{68,70}$Ni can be compared to those of $^{42,48}$Ca presented in Ref. \cite{LitvinovaSchuck2019}. 
Similarly, configurations $2q\otimes 2phonon$ included in REOM$^3$ induce a stronger fragmentation of the GDR and its additional spreading toward both higher and lower energies. Visible shifts of the main peaks of the giant dipole resonance (GDR) toward higher energies were obtained in $^{42,48}$Ca due to these high-complexity configurations. This observation was related to the appearance of the new higher-energy complex configurations and, consequently, the additional higher-energy poles in the resulting response functions. However, it was unclear whether or not the shift of the main peak is a generic feature of the $2q\otimes 2phonon$ configurations. The results for the $^{68,70}$Ni isotopes indicate that this effect is selective as no pronounced shifts are observed. As noted in \cite{LitvinovaSchuck2019}, the energy-weighted sum rule is conserved in REOM$^3$ because the analytical form of its dynamical kernel does not change as compared to REOM$^2$.
\begin{figure*}
\begin{center}
\vspace{-0.3cm}
\includegraphics[scale=0.52]{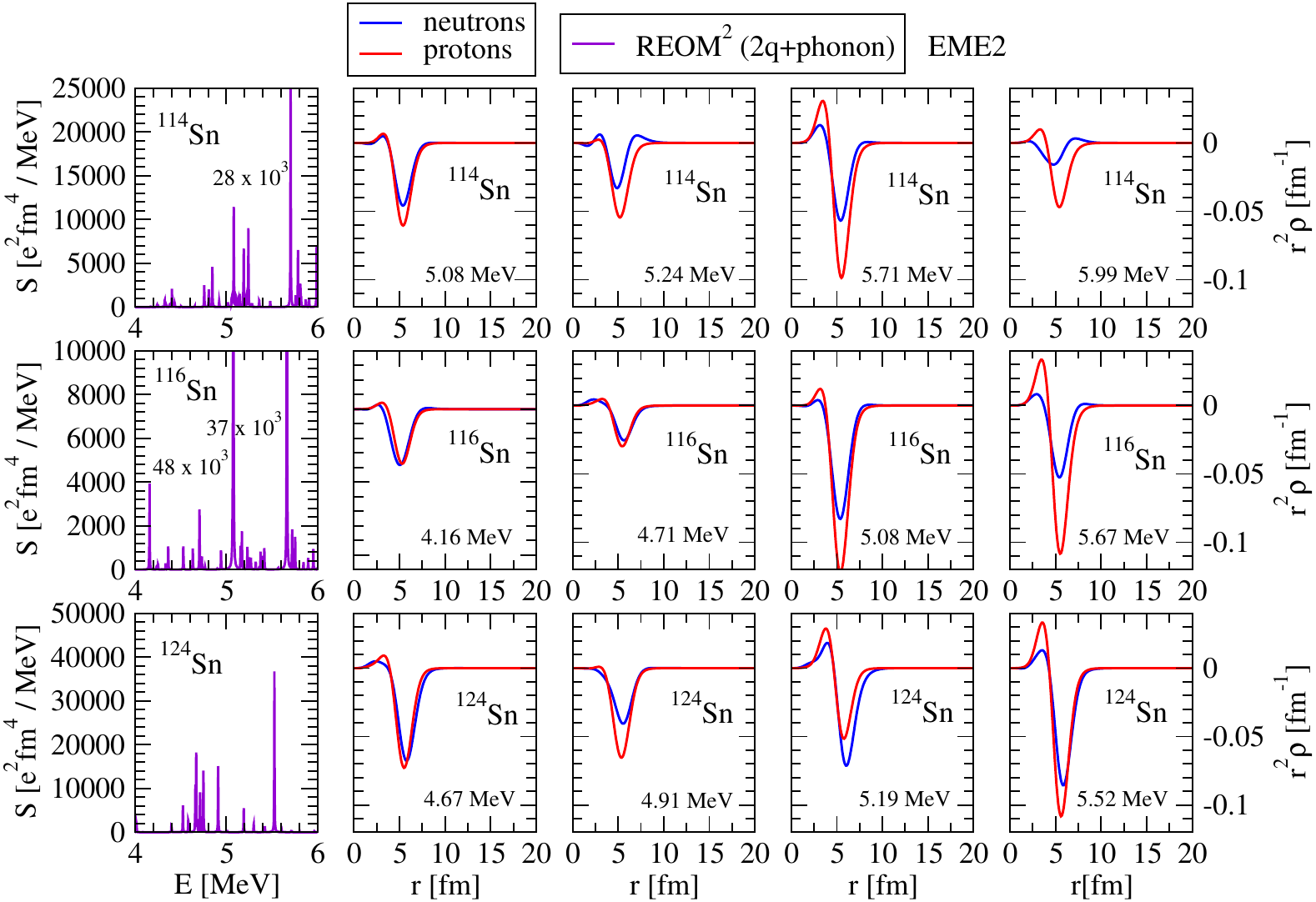}
\end{center}
\vspace{-0.6cm}
\caption{Same as in Fig. \ref{quad_1}, but for the quadrupole strength in the 4-6 MeV energy interval.}
\label{quad_2}%
\end{figure*}
%

The low-energy dipole response of the unstable neutron-rich $^{68,70}$Ni nuclei is of special significance as they are located on the r-process nucleosynthesis path. 
Because of its relevance to astrophysics, it has been investigated both experimentally \cite{Wieland2009,Rossi2013,Wieland2018} and theoretically \cite{LitvinovaRingTselyaev2010,LitvinovaRingTselyaev2013,Papakonstantinou2015}. 
The results of the RQRPA, REOM$^2$ and REOM$^3$ calculations for the low-energy dipole response of $^{68,70}$Ni are displayed in Fig. \ref{Ni-PDR} with the same curve and color-coding as in Fig. \ref{Ni-GDR}. In particular, one can see how richer spectra of REOM$^2$ and REOM$^3$ emerge from a relatively poor one of RQRPA. The latter is essentially the single strong and relatively collective state at $~9.5$ MeV in both nuclei with more strength in the case of $^{70}$Ni because of its larger neutron excess. The addition of $2q\otimes phonon$ configurations of REOM$^2$ results in the fragmentation of this state over a broader energy region. Finally, with the $2q\otimes 2phonon$ configurations, the fragmentation effect is further reinforced resulting in a distribution without a clear dominance of a single state but is rather spread uniformly over the $7-15$ MeV energy interval with a smooth strength increase toward the GDR. One can notice also the appearance of excited states at lower energies.  Thus, these examples illustrate how the three models with the increasing complexity of the dynamical kernel form a hierarchy that translates to the hierarchy of spectral functions with the increasing richness of their fine structure. The results in this energy region are compared to experimental data of Ref. \cite{Wieland2018}, where the two parts of the strength distribution in  $^{70}$Ni were obtained by different methods, that are indicated by different colors. One can see that a sequential increase of configuration complexity in the response theory improves the description, while the highest-complexity $2q\otimes 2phonon$ configurations still bring significant changes to the low-energy spectra. This suggests that such configurations are important for the quantities, which are sensitive to the details of the low-energy strength distributions, in particular, the radiative neutron capture rates in the r-process. Similar effects for the isospin-flip transitions and thus further refinements of the beta decay rates as compared to, e.g., Refs. \cite{RobinLitvinova2016,LitvinovaRobinWibowo2020} are expected. 
\begin{figure*}
\begin{center}
\vspace{-0.3cm}
\includegraphics[scale=0.52]{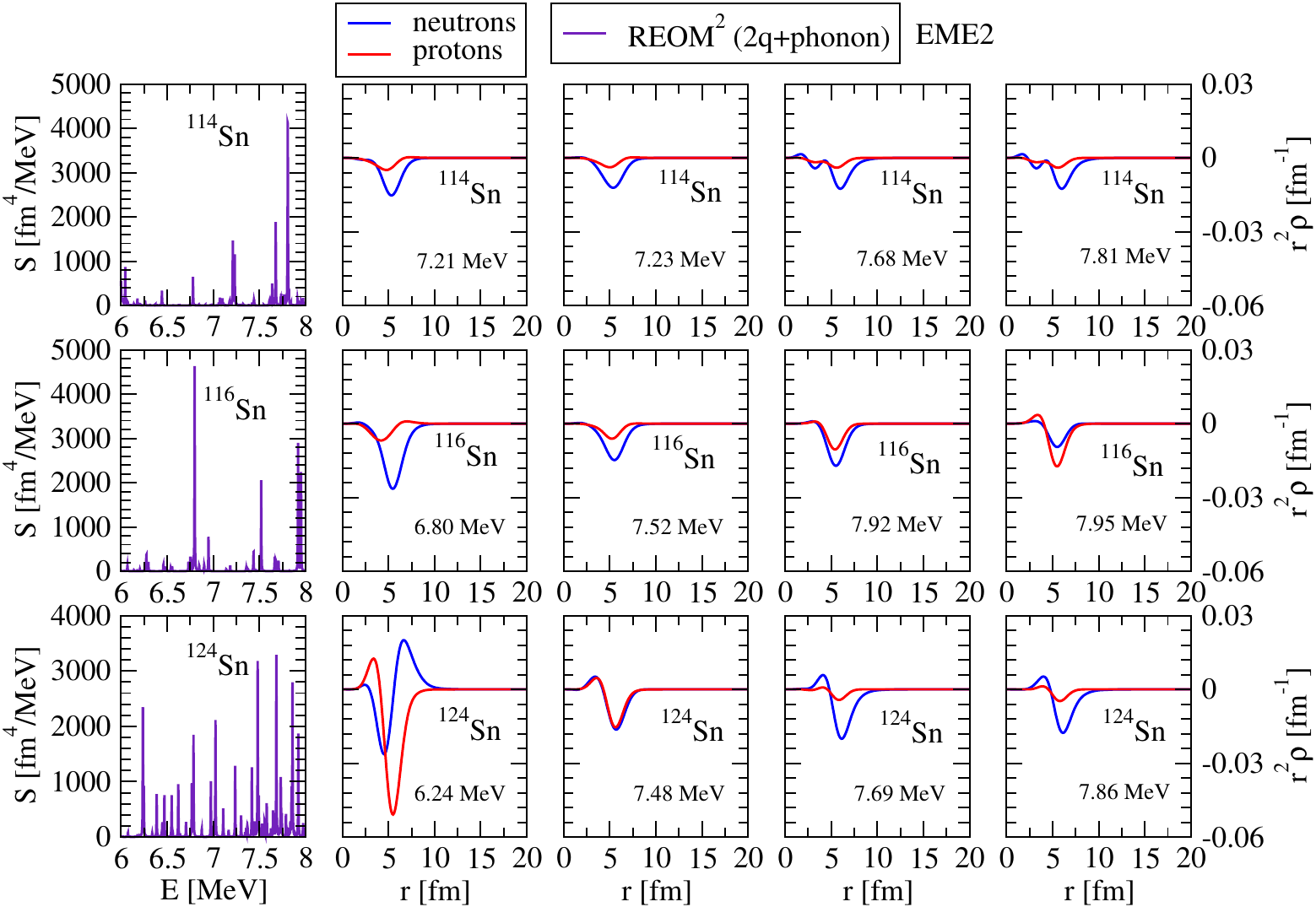}
\end{center}
\vspace{-0.6cm}
\caption{Same as in Fig. \ref{quad_1}, but for the quadrupole strength in the 6-8 MeV energy interval.}
\label{quad_3}%
\end{figure*}
%
%

The low-energy dipole strength, associated with the neutron skin oscillation and also known as pygmy dipole resonance (PDR), attracts active attention from both theory and experiment \cite{SavranAumannZilges2013,Lanza2023}.  Its fine structure and relevance for astrophysics were addressed in numerous RNFT studies \cite{Litvinova2007,LitvinovaRingTselyaevEtAl2009,LitvinovaLoensLangankeEtAl2009,EndresLitvinovaSavranEtAl2010,MassarczykSchwengnerDoenauEtAl2012,LitvinovaBelov2013,PoltoratskaFearickKrumbholzEtAl2014,Oezel-TashenovEndersLenskeEtAl2014,LanzaVitturiLitvinovaEtAl2014,NegiWiedekingLanzaEtAl2016}. The detailed knowledge on the low-energy strength of quadrupole character is limited, but also represents an interesting subject \cite{Tsoneva2011,Spieker2016,Tsoneva2019,Lenske:2019ubp}. In particular, in Ref. \cite{Tsoneva2011} the electric quadrupole response was investigated theoretically within the QPM along the Sn isotopic chain with special emphasis on excitations above the first collective state and below the particle emission threshold. This study has reported that depending on the asymmetry, a quadrupole strength clustering similar to the known PDR is found. The authors concluded from analyzing the transition densities of low-energy quadrupole states that the obtained results are compatible with oscillations of neutron or proton skins against the isospin-saturated core and with experimental data \cite{Spieker2016,Tsoneva2019}. 

Figs. \ref{quad_1} - \ref{quad_3} demonstrates the capability of RNFT to describe the quadrupole strength displaying the REOM$^2$ results for the 2$^+$ electromagnetic (EME2) excitations in three even-even tin isotopes $^{114,116,124}$Sn below 8 MeV, i.e.,  below their particle emission thresholds (10.3, 9.6, and 8.5 MeV, respectively). From the long tin isotopic chain, the $^{114,116}$Sn isotopes were selected for this study as the two typical mid-shell nuclei with similar moderate neutron excess, to investigate how the minimal difference in the neutron number translated to the differences and similarities in their 2$^+$ strength distributions. The $^{124}$Sn was chosen as the isotope with a significantly larger neutron number, potentially forming a prominent neutron skin. The structure of the quadrupole spectra in the considered energy region is qualitatively similar in all three nuclei: there is a strong collective lowest state at around 1 MeV and a group of states at higher energies with varying strength. The giant quadrupole resonance in these nuclei is located above 20 MeV and is not examined here. The RQRPA results for the lowest collective 2$^+$ state are given and discussed in Ref. \cite{AfanasjevLitvinova2015}. The major effect of adding $2q\otimes phonon$ configurations on this state is a minor downward shift, which brings its position closer to the experimental values, and a slight reduction of the transition probability. In the studies of Refs. \cite{Tsoneva2011,Spieker2016,Tsoneva2019} the lowest 2$^+$ state is interpreted as a phenomenon, which is structurally separated from the other quadrupole excitations at higher energies, while the latter ones were associated with pygmy quadrupole resonance. 

The REOM$^2$ EME2 spectra in $^{114,116,124}$Sn below 4 MeV is shown in the left panels of Fig. \ref{quad_1}, while the transition densities of the most pronounced excited states are given on the right-hand side of the respective strength distributions. The latter is calculated with a small value of the smearing parameter $\Delta = 2$ keV to resolve the fine structure of the strength. The reduced transition probabilities $B_n(E2)\uparrow$ can be related to the values of the strength functions at the peaks at $E = \omega_n$ as $B_n(E2)\uparrow = S_{\text{E2}}(\omega_n)\pi\Delta$. The transition densities of the lowest 2$^+$ state are prevailed by the in-phase surface oscillation of the neutron and proton Fermi liquids with some neutron dominance and in the case of $^{114,116}$Sn show out-of-phase overtones in the bulk. While moving toward the upper bound of the $0 \leq E \leq 4$ MeV energy interval, the next general trend is the weakening of the surface oscillations, which remain in phase, and the reduction of the neutron dominance. The next part of the spectrum illustrated in Fig. \ref{quad_2} in the same manner shows a significant increase of the density of states. The trends in the behavior of the transition densities of the strongest 2$^+$ excitations change showing the growth of the surface oscillation amplitudes and, interestingly, a proton dominance in all the nuclei for the majority of states in the  $4 \leq E \leq 6$ MeV energy interval. When moving beyond this energy interval, the density of states further increases and the transition densities change back to mainly neutron-dominant character, except for one state at E = 6.24 MeV in $^{124}$Sn. The reduction of amplitudes of the oscillations reflects stronger qPVC effects in the formation of the corresponding states in the $6 \leq E \leq 8$ MeV energy interval as the qPVC takes away considerable parts of the total normalization of the transition densities \cite{LitvinovaRingTselyaev2007}. Some enhancement in the surface oscillation intensity is noted in $^{124}$Sn that can be attributed to its neutron richness. At variance with Refs. \cite{Spieker2016,Lenske:2019ubp}, REOM$^2$ calculations do not show a dominance of out-of-phase oscillations of neutrons vs protons below 5 MeV. However, the transition densities for the 2$^+$ state at E = 6.24 MeV in $^{124}$Sn indicate that such oscillations can be present in the energy region under study. The near absence of such states, which are typical for the energies between the PDR and GDR, is consistent with the fact that the giant quadrupole resonance represents a $2\hbar\omega$ oscillation mode and therefore is located at a much higher energy than the GDR. This means, in particular, that the "pygmy quadrupole" mode may be spread over a wider energy interval toward higher energy, which calls for further detailed studies of this phenomenon.   


\section{Summary}
\label{summary}

 A theoretical framework for the low-rank fermionic propagators, most relevant to the observables in strongly-coupled superfluid fermionic many-body systems, is presented. The equations of motion for the two-times one-fermion and two-fermion superfluid propagators are obtained continuously with the only input of the bare two-fermion interaction. The EOM for the one-fermion propagator is worked out in the single-particle space and shown to simplify after the transformation to the HFB basis. This operation reveals important relationships between the two representations, scales the computational effort down considerably for realistic implementations, and paves the way to the superfluid response theory, which is obtained in the HFB basis from the start. 
 
The exact forms of the interaction kernels containing higher-rank propagators in their dynamical components are discussed. These propagators are approximated by factorizations into the possible products of two-fermion and one-fermion propagators in the superfluid regime, thereby introducing the truncation of the many-body problem on the two-body level, keeping the leading effects of emergent collectivity. It is thus and further shown how, with gradually relaxing correlations, the exact theory is reduced to the known approximations.

The major focus is then directed on the quasiparticle-vibration coupling in the dynamical kernels, where the normal and pairing phonons become components of the unified superfluid phonons. Self-consistent implementations are presented in the framework of the relativistic nuclear field theory for the dipole and quadrupole responses of the neutron-rich nickel and tin isotopes, respectively. The analysis of the obtained results is concentrated on the qPVC effects, which considerably modify 
the strength distributions, introducing their fragmentation and shifting the positions of the major peaks. Comparison of the strength distributions for the electromagnetic dipole response in $^{68,70}$Ni computed with configurations up to $2q\otimes 2phonon$ with experimental data indicates that increasing the configuration complexity of the dynamical kernel brings the theoretical results in better agreement with the data. At the same time, the resulting strength functions show saturation of its general features, so that the higher-complexity configurations may appear to be more important for the fine structure of the obtained spectra than for their gross structure. A study of the low-energy electromagnetic quadrupole strength was performed for the first time within the RNFT response theory confined by $2q\otimes phonon$ configurations for $^{114,116,124}$Sn and revealed some new insights in the underlying structure of the individual states forming this strength. 
The obtained results accentuate the importance of further advancing the quantum many-body theory in the sector of complex configurations which, in turn, give feedback on the static kernels of the fermionic EOMs. Addressing the latter aspect of the many-body problem is also necessary for heading toward spectroscopically accurate nuclear theory respecting special relativity and rooted in the standard model of particle physics.

\section*{Acknowledgement}
Illuminating discussions with Mark Spieker, Nadezhda Tsoneva, Enrico Vigezzi, and Horst Lenske are gratefully acknowledged.
This work was supported by the GANIL Visitor Program, US-NSF Grant PHY-2209376, and US-NSF Career Grant PHY-1654379.
%
\bibliography{Bibliography_Jul2023}
\end{document}